\documentclass{birkjour}
 \theoremstyle{definition}
 
 \theoremstyle{remark}

 \numberwithin{equation}{section}

\usepackage{tikz}

\usepackage{url}

\usepackage{breqn}

\newcommand{\be}{\begin{equation}}
\newcommand{\ee}{\end{equation}}
\newcommand{\barr}{\begin{array}}
\newcommand{\earr}{\end{array}}
\newcommand{\bea}{\begin{eqnarray}}
\newcommand{\eea}{\end{eqnarray}}
\newcommand{\beqa}{\be \begin{array}{rcl}}
\newcommand{\eeqa}{\end{array} \ee}

\newcommand{\ul}[1]{\underline{#1}}
\newcommand{\myol}[1]{\overline{#1}}

\newcommand{\dt}{{\cdot}}
\newcommand{\wdg}{{\wedge}}
\newcommand{\crs}{{\times}}

\newcommand{\half}{{\textstyle \frac{1}{2}}}

\newcommand{\etal}{\textit{et al.}}

\newcommand{\Om}{\Omega}

\newcommand{\cla}{{\mathcal{A}}}

\newcommand{\cld}{{\mathcal{D}}}

\newcommand{\clf}{{\mathcal{F}}}
\newcommand{\clg}{{\mathcal{G}}}

\newcommand{\clj}{{\mathcal{J}}}
\newcommand{\cll}{{\mathcal{L}}}

\newcommand{\clr}{{\mathcal{R}}}

\newcommand{\clw}{{\mathcal{W}}}

\newcommand{\da}{\partial_a}
\newcommand{\db}{\partial_b}

\newcommand{\gk}{\gamma_{3}}
\newcommand{\go}{\gamma_{0}}

\newcommand{\Rrev}{\tilde{R}}

\newcommand{\fo}{\myol{f}}

\newcommand{\ho}{\myol{h}}

\newcommand{\grad}{\nabla}

\usepackage{bm}

\usepackage{caption}
\usepackage{floatrow}%
\usepackage{subcaption}
\usepackage{wrapfig}

\usepackage{blkarray}

\newcommand{\blacktext}[1]{\textcolor{black}{#1}}

\newcommand{\kms}{{\rm\,km\,s^{-1}}}

\newcommand{\msun}{M_{\odot}}

\DeclareMathAlphabet{\mathbfit}{OT1}{cmr}{bx}{it}
\SetMathAlphabet\mathbfit{bold}{OT1}{cmr}{bx}{it}
\DeclareMathAlphabet{\mathbfss}{OT1}{cmss}{bx}{n}
\SetMathAlphabet\mathbfss{bold}{OT1}{cmss}{bx}{n}

\newcommand{\mpc}{{\rm \, Mpc}}

\newcommand{\hub}{\ul{\sfh}}

\newcommand{\hob}{\bar{\sfh}}

\newcommand{\fob}{\bar{\sff}}

\newcommand{\sft}{{\sf t}}
\newcommand{\sfx}{{\sf x}}
\newcommand{\sfy}{{\sf y}}
\newcommand{\sfz}{{\sf z}}

\newcommand{\oll}[1]{\overline{#1}}

\renewcommand{\fob}{\bar{f}}

\renewcommand{\hob}{\bar{h}}

\renewcommand{\hub}{\ul{h}}

\begin{document}

%
%
%
%
%
%
%
%
%

\title[Geometric Algebra, Gravity and Gravitational Waves]
 {Geometric Algebra, Gravity and Gravitational Waves}

\author[Anthony Lasenby]{Anthony N.\ Lasenby}

\address{%
Kavli Institute for Cosmology,\\
c/o Institute of Astronomy,\\
Madingley Road,\\
Cambridge CB3 0HA, UK
\\
and
\\
Astrophysics Group,\\
Cavendish Laboratory,\\
JJ Thomson Avenue,\\
Cambridge CB3 0HE, UK}

\email{a.n.lasenby@mrao.cam.ac.uk}



\date{today}

\begin{abstract}
\footnote{Sections 1 to 7 are based on parts of the plenary talks given at `AGACSE 2018: The 7th Conference on Applied Geometric Algebras in Computer Science and Engineering', July 2018, Campinas, Brazil and at `ICCA 11: The 11th International Conference on Clifford Algebras and Their Applications in Mathematical Physics', August 2017, Ghent, Belgium.}We discuss an approach to gravitational waves based on Geometric Algebra and Gauge Theory Gravity. After a brief introduction to Geometric Algebra (GA), we consider Gauge Theory Gravity, which uses symmetries expressed within the GA of flat spacetime to derive gravitational forces as the gauge forces corresponding to making these symmetries local. We then consider solutions for black holes and plane gravitational waves in this approach, noting the simplicity that GA affords in both writing the solutions, and checking some of their properties. We then go on to show that a preferred gauge emerges for gravitational plane waves, in which a `memory effect' corresponding to non-zero velocities left after the passage of the waves becomes clear, and the physical nature of this effect is demonstrated. In a final section we present the mathematical details of the gravitational wave treatment in GA, and link it with other approaches to exact waves in the literature. Even for those not reaching it via Geometric Algebra, we recommend that the general relativity metric-based version of the preferred gauge, the {\em Brinkmann metric}, be considered for use more widely by astrophysicists and others for the study of gravitational plane waves. These advantages are shown to extend to a treatment of joint gravitational and electromagnetic plane waves, and in a final subsection, we use the exact solutions found for particle motion in exact impulsive gravitational waves to discuss whether backward in time motion can be induced by strongly non-linear waves.
\end{abstract}

\maketitle
\section{Introduction\label{sec:intro}}

The past three years have seen a great deal of interest in gravitational waves, with their discovery at LIGO in early 2016. Gravitational waves are an outstanding example of the power of mathematical and physical theory to predict a new class of phenomenon which is only later verified by experiment. However, to most working physicists and engineers, general relativity and gravitational waves themselves seem a very difficult and complex area --- one where the mathematics is dominated by complex index manipulations and high level differential geometry, which only a few can confidently embark on and understand, and where the `physics' is full of non-intuitive elements, which make the nature of the real physical predictions of the theory difficult to pin down or grasp.

Indeed, in the case of gravitational waves themselves, while they were first discussed by Einstein in the context of general relativity (GR) in 1916, it took decades for their physical significance to be understood, and Einstein himself went through periods of doubting that they corresponded to anything physical. The case of black holes is similar, and it was perhaps even longer before an adequate understanding was reached as to whether they corresponded to something that might exist in the universe, and have physical effects. Thus even amongst professionals, GR is a difficult theory, for which the physical predictions can be difficult to understand and extract. It is therefore not at all surprising that amongst physicists, mathematicians and engineers working in other areas, there is an assumption that they will not be able to understand concepts such as gravitational waves or black holes properly, and that this problem concerns both the physics and mathematics involved.

What we wish to argue here, is that Geometric Algebra (GA) provides a route through to such understanding, and one which can reach much more widely (given an understanding of GA), than conventional approaches. By formulating general relativity as a gauge theory (similar to those of the strong and weak interactions) in flat space, written using the mathematics of GA, then the theory and the nature of its physical predictions become much clearer. This will be illustrated by the case of gravitational waves themselves, where the GA approach suggests a new `gauge' in which to study their physics, which has immediate and appealing links to electromagnetism, and which helps to iron out various misunderstandings and problems with gravitational waves and their detection which have surfaced before. Additionally, it clearly predicts a new type of `gravitational memory' effect, one which while it may be very small in most situations, nevertheless may have an interesting role to play in the paradoxes concerning information loss from black holes.

To start with this study of the role of Geometric Algebra in gravity, we will give a short survey of the basics of GA itself, highlighting those features that we have found to be particularly useful for studying gravity. We then follow this with a description of Gauge Theory Gravity, before considering solutions for black holes and gravitational waves in this approach. The features of solutions in the new gauge are discussed and some possibilities for their observation and for their theoretical relevance considered. This discussion is mainly in the nature of a review, rather than giving detailed mathematical derivations, but then in the final sections of this article we fill in many of the details, so that the nature of the new gauge and solutions can be clearly seen. This gauge has its parallel in metric-based General Relativity in something called the `Brinkmann metric', which while not widely known to astrophysicists, is argued here to be the preferred gauge in which to study gravitational plane waves, even for those not reaching it via Geometric Algebra.

Obviously in a contribution of this length it is not possible to give full details of either Geometric Algebra or Gauge Theory Gravity, so for those readers wanting a fuller account we refer to the book `Geometric Algebra for Physicists' \cite{d2003geometric} by Doran \& Lasenby, and the paper `Gravity, Gauge Theories and Geometric Algebra' by Lasenby, Doran \& Gull \cite{1998RSPTA.356..487L}. The recent review \cite{Lasenby:2016lfl} could also be useful, since it emphasises some different aspects of GA in gravity, and also contains a description of some applications of GA to electromagnetism, which is only treated very briefly here (in the context of joint EM and gravitational waves). Finally, we should note for those readers interested primarily in the particular `memory effect' for gravitational waves discussed here, that this has been independently discovered, at about the same time as the work reported here, and also related to the Brinkmann metric, by Gary Gibbons, Peter Horvathy and co-workers, and that the paper \cite{2017PhLB..772..743Z} would be good to consult on this, being the first in a series of papers by them on this topic.

\section{Geometric Algebra}

Geometric Algebra is a {\em covariant language} for doing physics and geometry. For two vectors $a$ and $b$, we can define the wedge and scalar products in terms of the Clifford (or {\em geometric}) product $ab$ via
\[
a\dt b = \half\left(ab+ba\right), \qquad a\wdg b = \half\left(ab-ba\right)
\]

Starting with a frame $\left\{e_i\right\}$, $i=1,\ldots,n$, we can then form the entire Clifford algebra. The $e_i$ are the {\em vectors}, $e_i \wdg e_j$ are the {\em bivectors} (grade-2 objects), $e_i \wdg e_j \wdg e_k$ are the {\em trivectors} (grade-3 objects), and so on, in the usual way up to
\[
e_1 \wdg e_2 \wdg \ldots \wdg e_n \propto I,
\]
where $I$ is the {\em pseudoscalar} for the space.
(Note that the generalised wedge product can be defined as the highest grade part of the geometric product between two objects with given grades.)

In geometry, Geometric Algebra is very good for doing {\em rotations} and {\em reflections}

\noindent\begin{minipage}{0.6\textwidth}
Let us start with {\em reflections:} quite generally, given a (normalised) object $B$ in the GA we can form a reflection in it of another object $A$ via
\[
A \mapsto \pm B A B
\]
E.g., suppose the object $A$ to be reflected is a vector $a$, and the object it is reflected in is the unit vector $n$, then
\[
\begin{aligned}
a \mapsto a' &= a - 2 a\dt n n,\\
&=a-(an+na)n \\
&\text{i.e.} \quad \boxed{a'= - nan}
\end{aligned}
\]
does what we want.
\end{minipage}
\begin{minipage}{0.5\textwidth}
\centering
\includegraphics[width=0.65\textwidth]{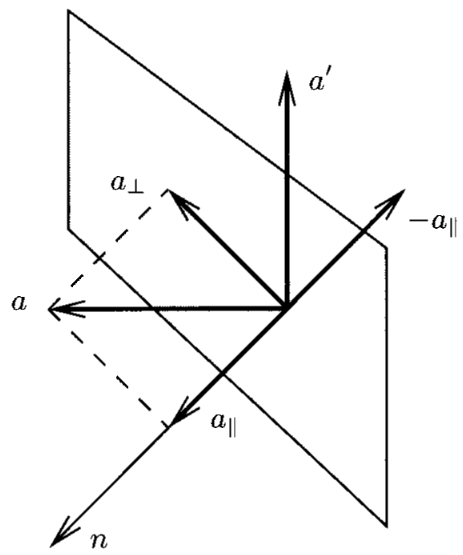}
\captionof{figure}{Reflection of a vector $a$ in a unit vector $n$.}
\end{minipage}

\noindent\begin{minipage}{0.5\textwidth}
For {\em rotations} we use the fact that a rotation in the plane generated by two unit vectors $m$ and $n$ is achieved by successive {\em reflections} in the planes perpendicular to $m$ and $n$.
To get from $a$ to $c$ we first form
\[
b=-mam
\]
and then perform a second reflection to obtain
\[
c=-nbn=-n(-mam)n=nmamn
\]
So if we define
\[
R= nm
\]
and the operation of {\em reversion}, which we indicate with a tilde, by
\end{minipage}
\begin{minipage}{0.5\textwidth}
\centering
\includegraphics[width=0.75\textwidth]{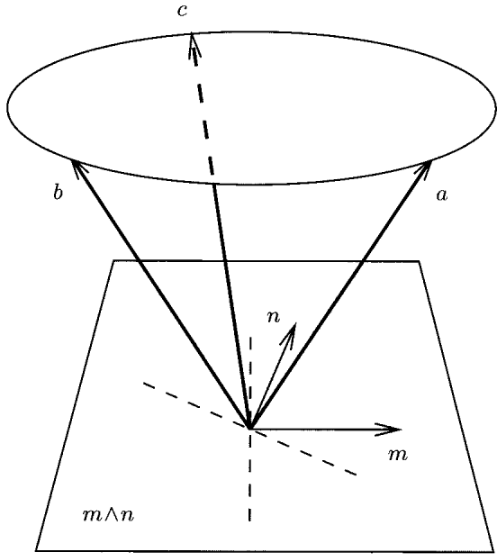}
\captionof{figure}{Rotation achieved via successive reflections.}
\end{minipage}
\[
\widetilde{\left(a b c \ldots p q\right)} = q p \ldots c b a \quad \text{\blacktext{for any set of vectors $a$, $b$, $c$ etc.}}
\]
then we can write the rotated vector as
\[
\boxed{c = R a \Rrev}, \quad \text{\blacktext{where we call}} \; R, \; \text{\blacktext{which satisfies}} \; R\Rrev=1 \; \text{{\em a rotor}}
\]

\subsection{Geometric Algebra as a language}

Now the key point, and what we meant by {\em covariant} above, is as follows.
Given geometric objects $A$, $B$, $C$, \ldots, we can form meaningful expressions by combining them with the `Clifford' or geometric product, generating $AB$, $ABC$, etc., and expressions derived from these like the wedge or dot products.

The expressions are meaningful if they are {\em covariant}, and this basically means that {\em if we carry out a transformation on each object individually, then this is the same as carrying out the transformation on the whole object}.

E.g., suppose we have a bivector $B= a \wdg b$, and rotate each of the vectors within it using a rotor $R$. Then
\[
\begin{gathered}
B \mapsto B' = R a\Rrev \wdg R b \Rrev = \half \left(R a\Rrev R b \Rrev - R b\Rrev R a \Rrev\right)= RB\Rrev \\
\text{\blacktext{i.e.}} \; \boxed{B'=RB\Rrev}
\end{gathered}
\]
Thus rotation using $R$ is a covariant operation, and we can confidently string vectors together in expressions knowing that the result is a geometric object transforming in the same way as the individual vectors.

The same applies to {\em reflections}, e.g. under reflections of $a$ and $b$ in a unit vector $n$ we have
\[
\begin{gathered}
B \mapsto B' = \left(-n an\right) \wdg \left(-n b n\right) = \half \left(n an n b n - n bn n a n\right)= nBn \\
\text{\blacktext{i.e.}} \; \boxed{B'=nBn}
\end{gathered}
\]

These two comments are what lies beneath the power of {\em Conformal Geometric Algebra} (CGA). Here we replace all conformal operations with either {\em rotors} (for translation, rotation and scaling), or {\em reflections} (for inversions). This gives us a powerful covariant language in which to express geometric relations.
 E.g.\, if the geometric object $S$ represents a sphere, and $A$ is another geometric object (which could be for example, a line, or a plane) then {\em inversion} of the object $A$ in $S$ is accomplished just by $A \mapsto \pm SAS$.

Now, we claim the same structure of geometric covariance underlies {\em gravity}.
(We will do this just in the usual structure of 4d-spacetime, but it is an interesting question of whether the CGA would be a better arena for this --- we will leave that for another day.)
To explain this properly, we will need two further aspects of GA --- {\em linear algebra} and {\em derivatives}.

\subsection{Geometric Algebra, Linear Algebra and Derivatives}

GA provides a beautiful framework for linear algebra --- the basic constructs are vector functions of vectors, e.g. $h(a)$ where this provides a vector for every input vector $a$, and is linear in the input, and
{\em Outermorphism} --- a powerful idea emphasised by David Hestenes, which extends $h$ to the entire algebra via (e.g.)
\[
h(a \wdg b) = h(a) \wdg h(b), \quad h(a \wdg b \wdg c) = h(a) \wdg h(b) \wdg h(c) \quad \text{etc.}
\]

The {\em adjoint function} $\ho(a)$ is defined (on vectors) by $a \dt h(b)=\ho(a)\dt b$. Simple but very non-trivial results in this approach are then
\[
\boxed{\det(h) = h(I) I^{-1}} \quad \text{and} \quad \boxed{\quad h^{-1}(A) = \det(h)^{-1} \ho(AI) I^{-1}}
\]
for the determinant, and for the inverse of $h$ on a general (homogeneous grade) object $A$.

For {\em derivatives}, there are three types of these.
Firstly, the standard Clifford differential operator $\grad$. Suppose we have some coordinates, $\left\{ x^\mu\right\}$, $\mu=0,1,2,3$ in spacetime, and a position vector $x$. Then we can define the frame of vectors $\left\{ e_\mu\right\}$ from them via $e_\mu = \frac{\partial x}{\partial x^\mu}$. Then we form the {\em reciprocal frame} $\left\{ e^\nu\right\}$, satisfying $e_\mu\dt e^\nu=\delta^\nu_\mu$.

We can then form the vector derivative
\[
\grad \equiv e^\mu \frac{\partial}{\partial x^\mu} \equiv e^\mu \partial_\mu
\]
(Note $e^\mu = \grad x^\mu$ is another way of thinking about this process.) The resulting object is then independent of the coordinates we started with. We can note also that for any vector field $a(x)$, the upstairs and downstairs components are just
\[
a^\mu \equiv a \dt e^\mu \quad \text{and} \quad a_\mu = a \dt e_\mu
\]
These statements look trivial, but are enough to do everything associated with vector calculus in curvilinear coordinate systems, which in standard expositions can look quite intimidating!

Secondly (introduced by {\em David Hestenes}), there is the {\em multivector derivative}.
We can only give a sketch of this, but the key starting quantity is the multivector derivative by a vector $a$, given, in a frame in which $a=a^\mu e_\mu$, by
\[
\da \equiv e^\mu \frac{\partial}{\partial a^\mu}
\]

For a general $n$-d space, and acting on a grade-$r$ object, these satisfy
\be
\begin{aligned}
\da a \dt A_r &= r A_r \\
\da a \wdg A_r &= (n-r) A_r
\label{eqn:da-res1}
\end{aligned}
\ee
\be
\text{\blacktext{and}} \quad \boxed{\da A_r a = (-1)^r (n-2r) A_r}
\label{eqn:da-res2}
\ee
Note the last of these means that if we differentiate a vector through a bivector, in 4d, the result vanishes. It is not obvious, but this turns out to be the key to why e.g.\ electromagnetism is a massless theory in 4d, and also being able to demonstrate how the {\em Riemann tensor} for a black hole works (see below).

Thirdly (and introduced by Lasenby, Doran \& Gull in \cite{Lasenby:1993ya}), one can extend this further to multivector derivatives with respect to a {\em linear function}, such as $h(a)$, not just a vector. If we write $h_{\mu\nu} = e_\mu \dt h(e_\nu)$, then we can assemble these into a frame-free derivative via
\[
\partial_{h(a)} \equiv a\dt e_\nu e_\mu \frac{\partial}{\partial h_{\mu\nu}}
\]
It is not expected to be obvious, but this is a wonderful tool in gravity, and means we can give coordinate- and index-free statements of all the main results and methods.

It is also very useful in linear algebra {\it per se}. E.g.\ here is a theorem which is quite hard to notate properly in a conventional matrix-based approach, but which we can write unambiguously and derive simply using the current approach:
\[
\partial_{h(a)} \det(h) = \det(h) \ho^{-1}(a)
\]
There are many other examples like this, and the power of the method has definitely not been fully explored yet.

\section{Gravity}

So finally we get to gravity! We want to consider a version of gravity that aims to be as much like our
    best descriptions of the other 3 forces of nature:
\begin{itemize}
    \item the {\em strong force} (nuclei forces)
    \item the {\em weak force} (e.g. radioactivity etc.)
    \item {\em electromagnetism}
\end{itemize}

These are all described in terms of {\em Yang-Mills type
    gauge theories} (unified in quantum chromodynamics) in a flat
    spacetime background. In the same way, {\em Gauge Theory Gravity (GTG)} is
    expressed in a flat spacetime. The key question is what we are gauging. We choose this to be {\em Lorentz rotations at a point}, and the ability to carry out an {\em arbitrary remapping} from one spacetime point to another. To motivate this, the Dirac equation and Dirac spinors are probably the easiest place to start, and so we now discuss these.

\subsection{Spinors in GA}

A key type of element in the GA is a {\em spinor}, which we can take for our purposes as a general even element of the algebra. So in 4d spacetime, one can write $\psi=$ scalar + bivector + pseudoscalar (8 d.o.f.)
and this is our version of a {\em Dirac spinor}.

It is helpful in discussing this to have a fixed frame of orthonormal vectors, $\left\{\gamma_\mu\right\}$, $\mu=0,1,2,3$ with $\gamma_0$ {\em timelike} ($\gamma_0^2=+1$) and the $\gamma_i$, $i=1,2,3$ {\em spacelike} ($\gamma_i^2=-1$). The Dirac equation is then
\[
\boxed{\grad \psi =- m \psi I \gk}
\label{eqn:dirac-eqn}
\]
which is quite simple!

Every Dirac wavefunction can be written in the form
\[
\psi \equiv \rho^{1/2}\exp\left( I\beta/2\right) R
\]
where $\rho$ and $\beta$ are scalars, and one soon finds that e.g.\ the Dirac current is $J= \rho v$, where the 4-velocity $v= R \gamma_0 \Rrev$. More generally we find
the following mapping of the $\{\gamma_\mu\}$ frame to a new frame which we can identify with some Dirac billinear observables.
\begin{center}
\includegraphics[width=.55\textwidth]{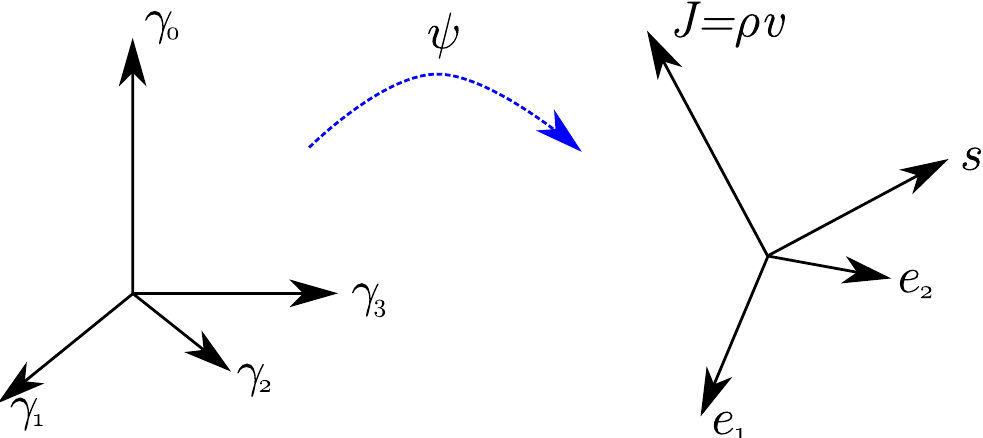}
\captionof{figure}{Action of $\psi$ on a fiducial frame.}
\end{center}
Here $s = \psi \gamma_3 \psi = \rho R \gamma_3 \tilde R$ is the spin vector, while
$e_1=\rho R \gamma_1 \tilde R $ and $e_2 =\rho R \gamma_2 \tilde R$ carry the phase information. This provides an interesting link between Dirac theory and the GA treatment of {\em rigid body mechanics}, where again one uses a rotor description to move between a fiducial set of fixed axes and moving axes accompanying the body (see e.g.\ Chapter~3 of \cite{d2003geometric}).

Because $\psi$ is (up to a pseudoscalar phase and scale) basically a rotor, if we carry out a further rotation of spacetime via a rotor $R'$ say, then $\psi$ responds single-sidedly
\[
\psi \mapsto R' \psi =\rho^{1/2}\exp\left( I\beta/2\right) R' R
\]
This explains the transformation law for spinors!

Note we can still combine spinors into covariant expressions but have to remember the single-sided transformation --- e.g.\ if $\psi$ is a spinor, and $v$ a vector, then $v\psi$ is a possible `phrase' of our covariant language (and transforms like a spinor), but $\psi v$ is not.

\section{Gauge Theory Gravity}

To motivate the transformations we consider in Gauge Theory Gravity (GTG), we start by considering two spinors (i.e.\ Dirac wavefunctions) $\psi_1(x)$
and $\psi_2(x)$.  A sample physical statement we might make within quantum mechanics is
\[
\psi_1(x) = \psi_2(x)
\]
i.e.\ at a point where one field has a particular value, the second field
has the same value.

This is {\em independent} of where we place the
fields in the STA. We could equally well introduce two
new fields
\[
\psi_1'(x) = \psi_1(x'), \quad \psi_2'(x) = \psi_2(x'),
\]
with $x'$ an arbitrary function of $x$.  The equation
$\psi_1'(x) = \psi_2'(x)$ has precisely the {\em same physical
content} as the original.

The same is true if we act on fields with a {\em spacetime rotor}
\[
\psi_1' = R\psi_1, \quad \psi_2' = R\psi_2
\]
Again, $\psi_1' = \psi_2'$ has same physical content as the original
equation. The only thing for which this does not work is {\em derivatives}.

For example, suppose $R$ in the rotation case is a function of position,
then
\[
\grad \left(R\psi\right) = \left(\grad R\right)\psi + \dot{\grad} R \dot{\psi} \neq R \left(\grad \psi\right)
\]
(here the dots indicate what the $\grad$ is operating on). We have failed to achieve a {\em covariant operation} in at least two ways --- firstly we have an inhomogeneous $\grad R$ term appearing, and secondly we have not managed to pass the vector derivative through $R$ in order to act directly on $\psi$.

Also position remapping will not work with derivatives, since if $x\mapsto f(x)$ (we call this a position gauge change), then it turns out that
\[
\grad_x \phi'(x) = \oll{f}\left(\grad_{x'}\phi(x')\right)
\]
where the linear function $\ul{f}(a)$ to which $\oll{f}$ is adjoint is given by $\ul{f}(a)=a\dt\grad f(x)$. I.e., an extraneous $\oll{f}$ gets in the way of covariance here.

We solve all these problems by introducing two gauge fields $\ho(a)$ and $\Omega(a)$. For $\ho(a)$, this is defined to have the transformation property $\ho(a) \mapsto \ho\left(\fo^{-1}(a)\right)$ under the position gauge change, so it is able to soak up the extraneous $\oll{f}$ if we use it to `protect' each derivative operator $\grad$, i.e.\ we henceforth use $\ho(\grad)$ instead of $\grad$.

For $\Omega(a)$, this allows Lorentz rotations (e.g.\ like $\psi \mapsto R\psi$) to be gauged locally (a rotation gauge change). The transformation property needed for this is
\[
\Om(a) \mapsto \Om'(a) = R \Om(a) \Rrev - 2 a \dt \grad R \Rrev
\]
The covariant derivative in the $a$ direction (for a quantity transforming double-sidedly) is
\[\cld_a\equiv a\dt\grad + \Omega(a)\crs\]
where the$\crs$ means the GA {\em commutator product}
\[
A\crs B\equiv \half(AB-BA)
\]
It turns out that the properties of the $\crs$ operator (basically, that it satisfies the Jacobi identity) together with the fact that $\Omega(a)$ is a bivector, mean that $\cld_a$ is a scalar operator and satisfies the Leibniz rule for derivatives.

We get a full vector covariant derivative via $\cld \equiv \ho(\partial_a)\cld_a$, where $\partial_a$ is the multivector derivative w.r.t. $a$ we discussed above.

The field strength tensor is obtained by commuting covariant derivatives:
\[
[\cld_a,\cld_b]M = R(a\wdg b) \crs M \qquad \text{($M$ \blacktext{ some multivector field})}
\]
This leads to the {\em Riemann tensor}
\[
R(a \wdg b) = a \dt \grad \Om(b) - b \dt \grad \Om(a) + \Om(a) \crs
\Om(b)
\]
from which we make a fully covariant version via $\clr(B)=Rh(B)$.
Note that geometrically, $\clr(B)$ is a mapping of {\em bivectors} to {\em bivectors}. (Also note that in \cite{Lasenby:2016lfl}, the expression for the Riemann given there (equation (5.5)) unfortunately contains two typographic errors --- the  $\da$ and $\db$ given there should have been $a \dt \grad$ and $b \dt \grad$, as here.)

The Ricci scalar is
\[\clr=\left(\partial_b\wdg\partial_a\right)\dt \clr(a\wdg b)\]
which is rotation gauge and position gauge invariant, and thus the simplest gravitational action to use is $\cll_{\rm grav}=\det h^{-1} \clr$, with the $\det h^{-1}$ being necessary to make the $d^4 x$ part of the action integral invariant.

The dynamical variables are $\ho(a)$ and $\Omega(a)$ and the field equations correspond to taking $\partial_{\ho(a)}$ and $\partial_{\Omega(a)}$. In the absence of matter the complete set of equations can be written in the useful form
\[
\boxed{\da \clr(a \wdg b) =0, \quad \cld\wdg \ho(a)=0}
\]
which are therefore relatively simple. All the symmetries of the Riemann that one encounters conventionally are encoded in the $\da \wdg \clr(a \wdg b) =0$ part of the first equation, and the second equation effectively says that the {\em torsion} vanishes in this case.

Further details and a full description of the general theory, including matter (which is allowed to have an intrinsic quantum spin) are contained in \cite{1998RSPTA.356..487L}, but what we have said so far provides enough detail for us to begin our discussion of black holes and gravitational waves. However, a few further comments about the nature of the resulting theory are in order.

Firstly, in terms of solutions, then locally the theory reproduces the predictions of an extension of General Relativity (GR)
known as {\em Einstein-Cartan} theory, which incorporates quantum spin and some possible torsion (of a restricted non-propagating form).
However, the current theory differs on global issues such as the nature of horizons, and topology (see \cite{1998RSPTA.356..487L}, particularly Section~6.4, for more details).

The advantages of GTG include being clear about what the physical
predictions of the theory are. Since it is a gauge theory, the physical predictions are the quantities that are {\em gauge-invariant!} Also it is conceptually simpler than standard GR, since it works in a flat space background. It is also simpler in a practical sense, since the covariant derivative  is implemented as a simple partial derivative plus the cross product with a bivector. This means that if one has a computer algebra program available that can do Clifford algebra in flat spacetime (for example, someone from an engineering background might well have this available, even when their focus hitherto has been on 3d Euclidean space, via a restriction of a 5d conformal geometric algebra program to 4d), then with such a program one can immediately start exploring {\em gravity}. In particular there is no need for getting familiar with a separate tensor calculus package, or indeed any need for consideration of curved space differential geometry.

Another advantage is that this approach also articulates very well with the {\em Dirac equation}. We can incorporate the effects of gravity into the flat space free-particle equation (\ref{eqn:dirac-eqn}) by promoting $\grad$ to $D$, where $D$ is the version of $\cld$ appropriate to spinors, namely
\be
D\psi \equiv \ho(\da)D_a \psi, \quad \text{where} \quad D_a = a\dt\grad + \half \Omega(a)
\ee
Again the only objects we need available if we wish to do computations for the Dirac equation in a gravitational field, are the elements of the STA already introduced, in which $\psi$ is a general even element, and we are able to work in a flat space background. This leads to conceptual simplifications which enabled (for example) the first computations of the spectrum of a fermion in a spherically symmetric gravitational potential (the analogue for a black hole of the Balmer series for an atom), in \cite{2005PhRvD..72j5014L}.

As a final general point, it is worth considering a further aspect of the novelty of our gauge theory approach, and one on which I had discussions with Waldyr Rodriguez, before his untimely death.

The covariant derivative $\cld=\ho(\da)\left(a\dt\grad + \Omega(a)\crs\right)$ cannot be taken as being the same as the conventional covariant derivative $\grad_\mu$. First of all, $\cld$ is a Clifford operator that is applied to other Clifford algebra geometric objects using the standard rules of flat-space geometric algebra. So, e.g., as already stated, if you have a computer algebra program that can do Clifford algebra in spacetime, then you can immediately start exploring {\em gravity}.

This contrasts with the conventional covariant derivative $\grad_\mu$, which is an abstract object that you will need the machinery of a full tensor calculus package to be able to work with.

But most importantly, the space that our covariant vectors live in simply does not exist in any conventional treatments of differential geometry.
In the following picture, the $\partial_\lambda x$ part indicates the conventional {\em tangent space} and the $\grad \phi$ part indicates the conventional {\em co-form} or {\em 1-form space}.
\begin{center}
\includegraphics[width=0.7\textwidth]{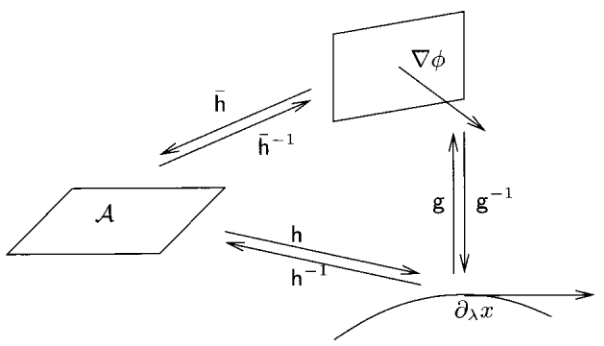}
\captionof{figure}{The spaces of conventional differential geometry (top and bottom), compared to the space of covariant objects in Gauge Theory Gravity (middle).}
\end{center}
In our approach we use the $\ho$-field and its transpose and inverses to make all
vectors of the same type --- covariant vectors, which live in the space marked $\cla$ (since the covariant form of the electromagnetic potential, $\cla$, is a typical example of such a vector). Then we just have rotor group
transformations which act within this space, which as stated is not available in conventional approaches.

$g$ here is our version of the metric tensor, which conventionally (and here) maps between the {\em tangent} (vector) and {\em cotangent} (1-form) spaces.
We can write
\[
g= \hob^{-1} h^{-1}
\]
and in components recover the standard GR metric as
\[
g_{\mu \nu} =  h^{-1}(e_\mu) \dt  h^{-1}(e_\nu)
\]
but in fact we never have any need to do this! In practice it is better to work in terms of the $\hob$ function.

There is probably a lot more to explore in relation to the rest of differential geometry, not least of course the fact that everything we have been doing here in gravity is in a {\em flat space!}


\section{Black holes}

Two very current aspects of general relativity are black holes and gravitational waves, linked in the first detection of gravitational waves by the LIGO interferometers (see Fig.~\ref{fig:bh-sim} for the simulated appearance of these black holes). Here we wish to discuss how these two central parts of GR look in Gauge Theory Gravity, starting with black holes.

\begin{minipage}{0.45\textwidth}
Just like setting up the EM equations for a point charge, we need to choose a {\em gauge} and work from there. We would like a gauge (choice of $\ho$-function) that covers all of (flat) space, except possibly a singularity at the origin.
Note that, again just like EM, we would expect the field strength tensor to be independent of our choice of gauge.
\end{minipage}
\begin{minipage}{0.55\textwidth}
\centering
\includegraphics[width=0.725\textwidth]{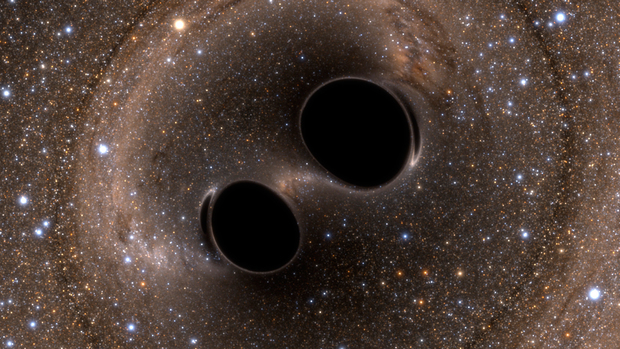}
\captionsetup{width=0.725\textwidth}
\captionof{figure}{Simulation of the binary black hole pair responsible for the first gravitational wave detection (credit: SXS group).\label{fig:bh-sim}}
\end{minipage}

\vspace*{0.1in}

Denoting $e_r$ as the unit radial vector, $e_t=\go$ as the unit time vector and the radial null vector $e_-=e_t-e_r$, then two good choices for $\ho$ are the following:
\[
\begin{aligned}
\ho(a) &= a -\sqrt{\frac{2M}{r}} (a \dt e_r) e_t \\
\text{\blacktext{and}} \qquad \ho(a) &= a +\frac{M}{r} (a \dt e_-) e_-
\end{aligned}
\]
We call the first the {\em Newtonian gauge} since a lot of the physics looks very Newtonian-like in this gauge, and the second is the GTG analogue of the {\em Advanced Eddington-Finkelstein metric} (which is good for treating the motion of photons).

Both are pretty simple!
They both lead to the same Riemann tensor
\[
\clr(B) = -\frac{M}{2 r^3} \left(B+3 \sigma_r B \sigma_r\right)
\]
where $\sigma_r= e_r e_t$ is the unit spatial bivector in the radial direction.

We can immediately check the field equation $\da \clr\left(a \wdg b\right)=0$ is satisfied. Using the results for the $\da$ derivative above, in equations (\ref{eqn:da-res1}) and (\ref{eqn:da-res2}), we have
\[
\da\left(a\wdg b+3 \sigma_r (a \wdg b) \sigma_r\right) = 3 b +3 \da \left(\sigma_r (ab-a\dt b) \sigma_r\right)
=3b-3b \sigma_r^2=0
\]
where the result that differentiating a vector through a bivector gives zero in 4d (here $\dot{\da} \sigma_r \dot{a}=0$), is a crucial step.
This is quite impressive as regards compactness and ease of working.
Even more impressive is doing the same for a rotating black hole --- the {\em Kerr} solution, which we now consider.

\subsection{Rotating black holes}

Here if the black hole has angular momentum parameter $L$, we find
\[
\clr(B) = -\frac{M}{2 \left(r+IL\cos\theta\right)^3} \left(B+3 \sigma_r B \sigma_r\right)
\]
i.e.\ we get to this from the Schwarzschild (non-rotating) black hole via $r \mapsto r+IL\cos\theta$. This explains the {\em complex structure} previously noticed in the Kerr solution (e.g.\ \cite{Newman:1965tw}), but in terms of the spacetime pseudoscalar $I$, rather than an uninterpreted scalar imaginary $i$.

Notice we do not need to do any more work to show that $\da \clr\left(a \wdg b\right)=0$ is satisfied --- it follows from what we did in the Schwarzschild case, since $\da I =-I \da$. Of course quite a lot of work is necessary to get from an $\ho$-function to the Riemann in this case, but this is certainly the most compact form of Riemann for the Kerr in the literature (most authors do not even try to write down the Riemann components!).

As regards the $\ho$-function itself, using GA methods, Chris Doran was able to find a compact $\ho$-function gauge for the Kerr which is similar to the Newtonian gauge form for Schwarzschild --- the metric form of this is known as the {\em Doran metric} --- see \cite{2000PhRvD..61f7503D}. This uses {\em oblate coordinates}, and is therefore not as simple to describe as the Schwarzschild Newtonian form of $\ho$, but promises a similar simplicity of description of the physics  of infalling material as in the Schwarzschild case.


\section{Gravitational waves}

We now get to the central topic of this contribution, {\em gravitational waves}.

\begin{figure}
\begin{center}
\includegraphics[width=0.8\textwidth]{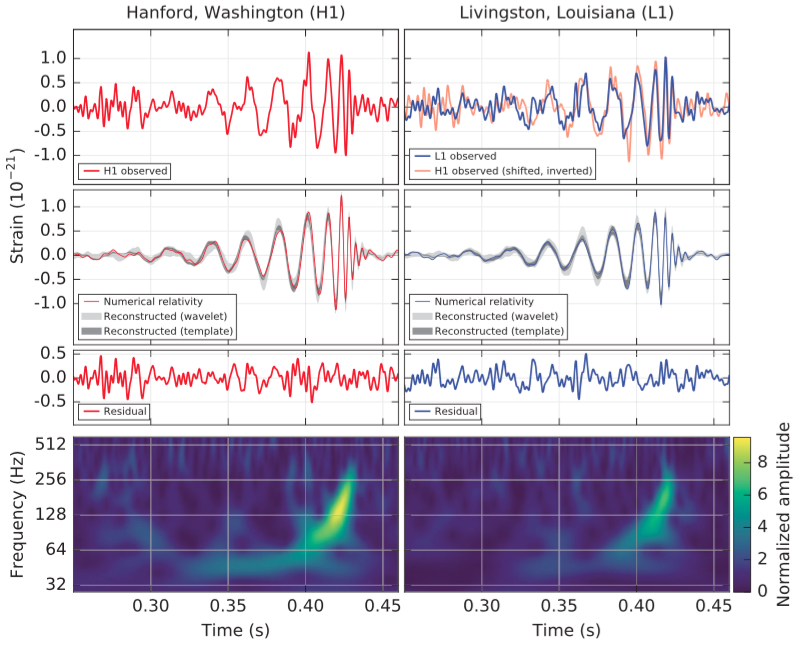}
\caption{The detection plots for the first detection of gravitational waves, made by the LIGO observatories in the US, and reported in Abbott \etal, \cite{2016PhRvL.116f1102A}.}
\label{fig:GW-detections}
\end{center}
\end{figure}

As mentioned above, and illustrated in Fig.~\ref{fig:bh-sim}, the first detection of gravitational waves, made in September 2015, was of the final stages of coalescence of two black holes, each with mass about $30\msun$. These were detected by the two interferometric observatories, one at Livingston, Louisiana, and the other at Hanford, Washington State, which make up the Advanced LIGO detector in the United States.
Fig.~\ref{fig:GW-detections}
shows the plots of `strain' (which we define below), and frequency of oscillation versus time observed at the two detectors. The sudden increase of frequency towards the end of the traces is called the `chirp' phase, and indicates where the gravitational wave energy radiated is sufficient to make the previously roughly circular orbits of the two black holes turn into steep spirals, at the end of which the black holes actually coalesce, producing a final black hole which no longer radiates.

The initially spherical gravitational waves (GWs) produced by the black holes will appear plane to distant observers, and we discuss the conventional approach to these below. First, however, we look at how we can represent plane GWs within Gauge Theory Gravity.

\subsection{The GTG approach to gravitational waves}

Some early versions of gravitational waves in the GTG approach were contained in the 1998 paper (\cite{1998RSPTA.356..487L}), which looked at some forms for the Riemann tensor for such waves, and their place in what is called the {\em Petrov classification}. Recently, I have been looking at them again, from the physical point of view, and particularly their effects on particles as they pass over them.

It was natural for me to start with a plane analogue of the Advanced Eddington Finklestein $\ho$-function for black holes discussed above: $\ho(a) = a +\frac{M}{r} (a \dt e_-) e_-$, where $e_-=e_t-e_r$. This is in what (in metric terms) is called a {\em Kerr-Schild} form, so I wanted a Kerr-Schild form for the planar case, which in rectangular coordinates, and for a wave propagating in the $z$ direction, would look like
 \be
 \boxed{\ho(a)=a-\half H \, a\dt e_+ \, e_+}
\label{eqn:h-KS}
\ee
where $e_+=e_t+e_z$, and $H=H(t,x,y,z)$ is a scalar function of spacetime position.

This would be a natural choice, since in the same way the Advanced Eddington Finklestein gauge is good for treating the motion of massless particles (photons), one might hope that its planar analogue would be good for treating gravitational waves themselves, which (in particle terms) are also massless.

It turns out that the form (\ref{eqn:h-KS}) works very well indeed, and
a remarkable feature is that despite being very simple, it provides an {\em exact} solution for gravitational waves. Further details are given below, but it turns out that with the ansatz $H(t,x,y,z)=G(\eta) f(x,y)$, where $\eta\equiv t-z$, one finds that $\da \clr\left(a \wdg b\right)=0$ is satisfied provided the 2d Laplacian $\grad^2 f=0$.

Using polar coordinates $(\rho,\phi)$ for the 2d $(x,y)$ plane, the solutions of $\grad^2 f=0$ that are picked out as giving homogenous values for the Riemann (i.e.\ the same all over the plane wavefront) are
\[
f=\rho^2\cos2\phi \quad \text{and} \quad f=\rho^2\sin2\phi
\]
and borrowing some freedom from $G(\eta)$, we get the final form of Riemann:
\[
\boxed{\clr(B)=\half G(\eta) \, (e_+ e_{\perp}) B (e_+e_{\perp})}
 \]
where $e_{\perp}=\cos(\phi_0(\eta))e_x+\sin(\phi_0(\eta))e_y$ is the arbitrary polarization direction in the $(x,y)$ plane. This is very neat in showing us how the input bivector $B$ is reflected in the bivector $e_+ e_{\perp}$, which encodes both the direction of propagation in spacetime, and the direction of polarization. The way that the polarization angle is given by $\phi_0$, whereas the solution for $H$ and the components of the Riemann rotate through $2\phi_0$ (see Section~\ref{eqn:sect-exact-waves}, for more details on this) is a consequence of the {\em `spin-2'} nature of gravitational radiation, and it is interesting to see it arising here due to the fact we are {\em reflecting} in the polarization direction.

Notice also how simple it is to see that the field equation $\da \clr\left(a \wdg b\right)=0$ is satisfied. The `pulse' $G(\eta)$ is just a scalar term, so we need
\[
\begin{aligned}
\da\left(e_+ e_{\perp} (a\wdg b) e_{\perp} e_+\right)&=\da\left(e_+ e_{\perp} (ab-a\dt b) e_{\perp} e_+\right)\\
&= -b e_+ e_{\perp} e_{\perp} e_+ =  b e_+ e_+=0
\end{aligned}
\]
which follows since differentiation through a bivector yields 0, and $e_+$ is null.

\subsection{Comparison with conventional approach}

So how does our version of gravitational waves compare with the conventional approach? The effects of gravitational waves are usually treated using what's called the $TT$ {\em (transverse traceless)} metric. Here the Einstein equations have been linearised, and for a wave going in the $z$-direction
we change the metric entries in the $x$ and $y$ directions, leaving the $z$ and $t$ directions alone. Specifically, the linearisation consists of writing the metric as
\[
g_{\mu\nu} = \eta_{\mu\nu}+h_{\mu\nu}
\]
where $\eta_{\mu\nu}$ is the Minkowski space (i.e. special relativity) metric, and we assume the perturbations $h_{\mu\nu}$ satisfy $h_{\mu\nu} \ll 1$. The entries in $h_{\mu\nu}$ are basically the `strains' referred to above, and for example in the first gravitational wave detection, shown in Fig.~\ref{fig:GW-detections}, are of the order of $10^{-21}$. (For reference, the gravitational field at the surface of the Earth corresponds to a strain of about $10^{-9}$.) We then form the `trace reversed' version of the $h_{\mu\nu}$ given by
\[
\bar{h}^{\mu\nu} = h_{\mu\nu} - \half \eta_{\mu\nu} h
\]
where $h=h^\sigma_\sigma$, in which the equations take their simplest form. The $TT$ gauge solution for a wave moving in the $z$ direction is then
\[
\bar{h}^{\mu\nu} = A^{\mu\nu} \exp(ik_\rho x^\rho),
\]
with $k^\mu=(k,0,0,k)$ and
\[
A^{\mu\nu} =
\begin{blockarray}{ccccc}
\begin{block}{*{5}{>{$\footnotesize}c<{$}}}
      & $t$ & $x$ & $y$ & $z$ \\
\end{block}
\begin{block}{>{$\footnotesize}c<{$}[*{4}{c}]}
  $t$ & 0 & 0 & 0 & 0 \\
  $x$ & 0 & a^+ & a^\times & 0 \\
  $y$ & 0 & a^\times & -a^+&  0 \\
  $z$ & 0 & 0 & 0 & 0 \\
\end{block}
\end{blockarray}
 \]
Here $a^+$ and $a^\times$ are in general complex numbers, and we have labelled the rows and columns of the $A$ matrix so that it is clear which directions are affected.

This is effectively the opposite of what we are doing in the GTG approach --- constructing a metric from our $h$-function, one finds that it has non-zero changes in the $z$ and $t$ directions and leaves the $x$ and $y$ directions alone. Specifically, using the $H = G(t-z) f(x,y)$ (where $G(t-z)=G(\eta)$) introduced above, we have
\[
h_{\mu\nu} =
\begin{blockarray}{ccccc}
\begin{block}{*{5}{>{$\footnotesize}c<{$}}}
      & $t$ & $x$ & $y$ & $z$ \\
\end{block}
\begin{block}{>{$\footnotesize}c<{$}[*{4}{c}]}
  $t$ & H & 0 & 0 & -H \\
  $x$ & 0 & 0 & 0 & 0 \\
  $y$ & 0 & 0 & 0 &  0 \\
  $z$ & -H & 0 & 0 & H \\
\end{block}
\end{blockarray}
\]

Note in this case we do not need to use the trace-reversed $h_{\mu\nu}$'s, and the sum, $g_{\mu\nu} = \eta_{\mu\nu}+h_{\mu\nu}$, provides a solution to the {\em exact} equations.

\begin{figure}
\begin{center}
\includegraphics[width=0.8\textwidth]{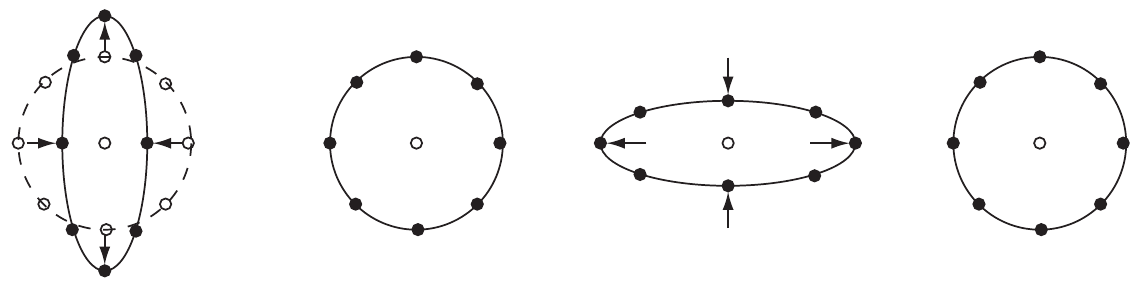}
\caption{Effects of the passage of a gravitational wave travelling in the $z$-direction on a ring of particles in the $xy$-plane (from \cite{2006gere.book.....H}).}
\label{fig:squeezing}
\end{center}
\end{figure}

Leaving aside the linearisation, which is more sensible?
A key step is to look at the {\em effect of the passage of the wave on particles in its path}. In the standard approach, in the $TT$ gauge, one often sees diagrams such as that in Fig.~\ref{fig:squeezing},
which is showing the effects on a ring of particles in the $(x,y)$ plane. However, calculating the geodesic equations in this ($TT$ metric) case, one finds that there is actually {\em no force} on these particles! Particles initially at rest in the $(x,y)$ plane, remain at rest. What is being indicated then, are changes in the {\em proper distances} between the particles, due to the changing geometry.

In the GTG approach, the geodesic equations are replaced by
\[
v \dt \cld v = \frac{d v}{ d s} + \omega(v) \dt v = 0
\]
where $v$ is the 4-velocity of the particle of interest, $s$ proper time along the path, and $\omega(a)\equiv \Omega h(a)$ is the position gauge covariant form of the $\Omega$ bivector gauge field. With our gauge choice for $h$ of the Kerr-Schild form above, this then leads to {\em explicit forces}, $-\omega(v)\dt v$ on the particles in the $(x,y)$ plane, and the effects on particles as depicted in the alternate squeezing in two directions, become actual motions of the particles.

So which is `right'? One might think either approach is valid --- all that matters is what we predict for physically observable quantities. However, there exists a subtlety. It turns out the linearisation used in the $TT$ gauge removes an effect I now think is important. This is the {\em net velocity imparted to the particle by the passage of the wave}.

\begin{figure}
\includegraphics[height=0.9\textheight]{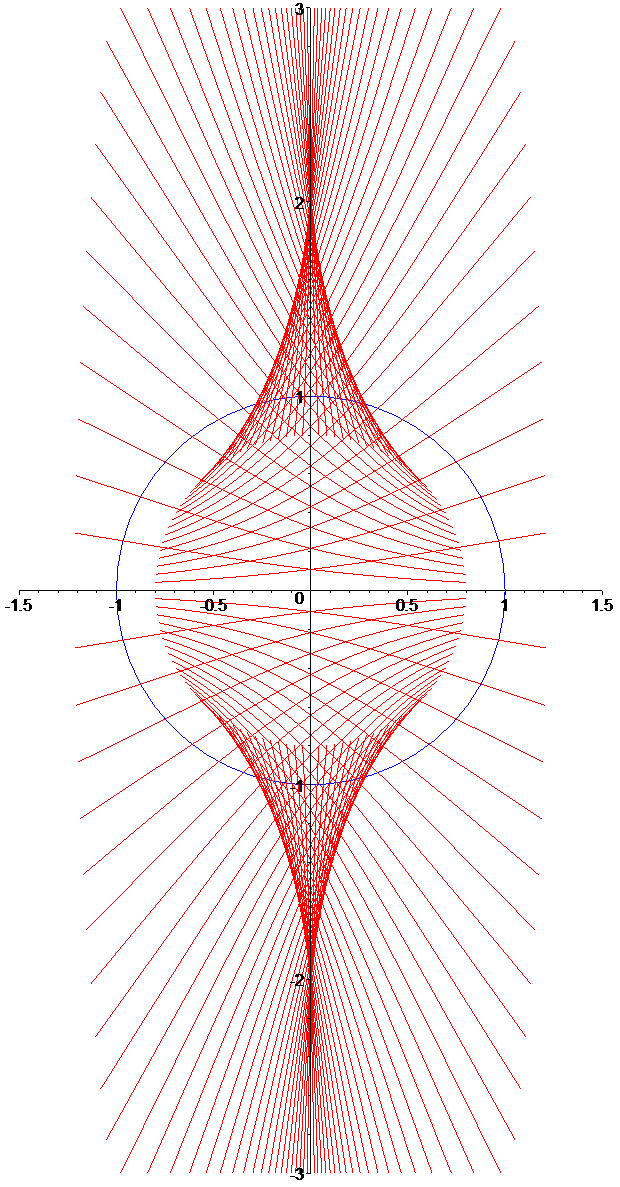}
\caption{Caustic formation in dust induced by the passage of a gravitational wave. A ring of particles is initially stationary in the $xy$-plane. (The ring has radius 0.8, with the blue circle of radius 1.0 being shown as a guide.) The wave is travelling into the page, in the $z$-direction, and the particles acquire a non-zero velocity after the wave has passed.}
\label{fig:2dcaust}
\end{figure}

\begin{figure}
\includegraphics[width=\textwidth]{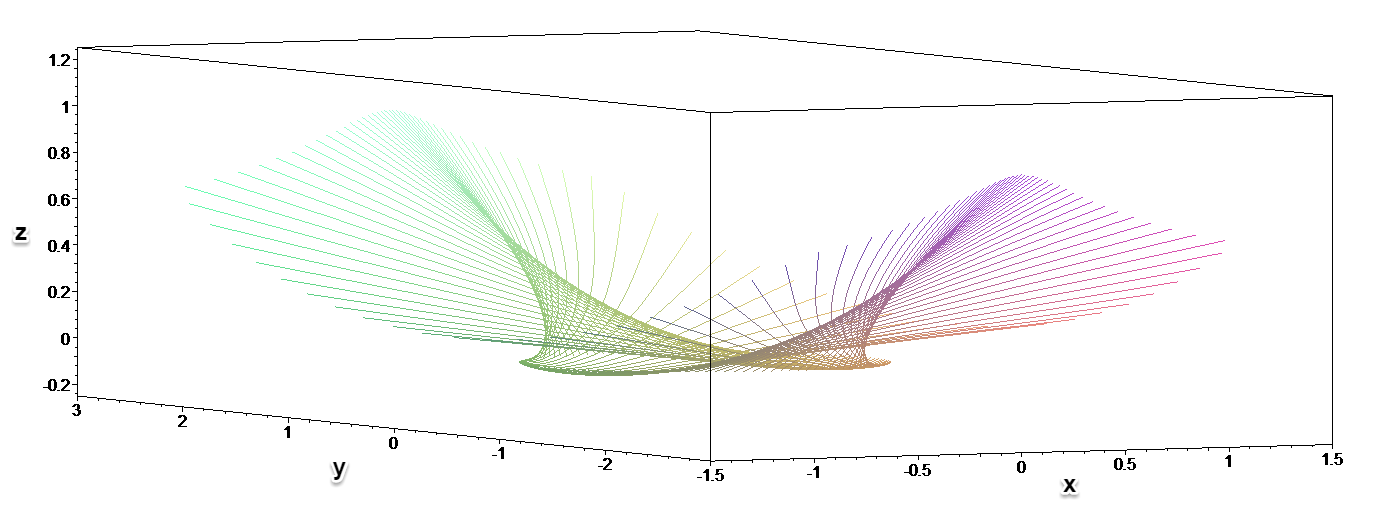}
\caption{3d view of the situation shown in Fig.~\ref{fig:2dcaust}.}
\label{fig:3dcaust}
\end{figure}

In our approach, one finds that the wave imparts a net velocity to the test particle that persists after the wave has passed. The direction of motion depends on initial position in the $(x,y)$ plane versus polarization angle. One gets some rather beautiful patterns, including the formation of caustics, as seen in Figures~\ref{fig:2dcaust} and \ref{fig:3dcaust} (more details of the particle motions shown in the Figures, and of how they were computed, are given in Section~\ref{sect:part-motion} below).

These `velocity memory' effects are entirely absent in the standard $TT$ approach --- the linearisation loses them. So could we recover them by seeking an exact version of the $TT$ gauge?
Historically, this was actually the route first explored for exact gravitational waves.

\begin{wrapfigure}{R}{0.375\textwidth}
\centering
\includegraphics[width=0.9\textwidth]{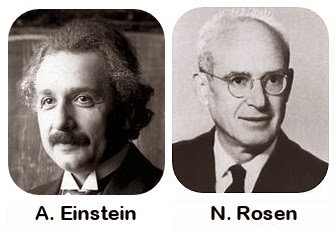}
\caption{Pictures of Einstein and Rosen, who were the first to examine exact gravitational waves (see \cite{1937FrInJ.223...43E}).}
\end{wrapfigure}

It is well known that in the 1930s, Einstein and Rosen attempted to work out {\em exact} solutions for gravitational waves in GR (the solutions to this point had been linear approximations), and found that apparently every wave was accompanied by an unphysical contraction of all space to one point following its passage. Because of this,
Einstein temporarily gave up believing that gravitational waves existed at all --- he only started believing again once it was established that the `collapse' was a type of coordinate singularity --- not physical. (It is interesting that Rosen never believed in GWs again!)

The waves they were studying were exact versions of the $TT$ gauge, which again had no forces on the particles --- just changes of proper distance. I now believe that the `contraction of all space to a point' was the exact $TT$-gauge's version of (at least some of) the particles being deflected, and approaching the origin after the passage of the wave.
So is this observable? One can estimate (see below) that the velocity deflection for a pair of particles is roughly
\be
\Delta v \sim - x_0 \int G \, d\eta
\label{eqn:effect-est}
\ee
where $x_0$ is their initial separation and a $\phi_0=0$ polarized wave is assumed. The sign of velocity deflection is opposite in the $y$ direction, and with equal magnitude. Also note $G(\eta)$ is a Riemann tensor eigenvalue, and hence an intrinsic quantity. We now provide some estimates of this effect in relevant astrophysical circumstances.

The most plausible scenarios involve binary black hole systems with much larger mass than those that have been detected by LIGO. This is because we want a strain as large as possible to get the biggest possible effect (the relation between `strain' and $G(\eta)$ is discussed in Section~\ref{sect:back-curv-and-em} below). We can move up to strains in the region of $10^{-13}$ or even larger by considering black hole masses appropriate to those known to exist at the centre of galaxies, which lie in the range of a few times $10^6\msun$, such as the black hole in the centre of our own Milky Way galaxy, up to of order $10^{10}\msun$ in massive galaxies. Mergers and collisions between galaxies are quite frequent, and when they occur the black holes in their centres may end up in orbit about each other, forming a supermassive binary pair. These will gradually lose energy by gravitational radiation, very much like a scaled up, and slowed down, version of the binary pair of black holes which led to the first detection. The frequencies of waves emitted by such a pair as they gradually inspiral, are much too low (in the range nanoHz to microHz!) to be detected by Earth-bourne interferometers such as LIGO. However, they can plausibly be detected from space, using either existing astronomical objects, or special spacecraft we place there for the purpose. In the first category, I have recently been involved in a proposal to use the apparent motions of stars visible in the survey of several billion star positions currently being carried out by the Gaia satellite, as a means of detecting such ultralow frequency waves --- see \cite{Moore:2017ity} for details. It seems unlikely that we could detect `velocity memory' in such a manner, since just detecting the waves themselves is at the limits of sensitivity, and the continuous nature of the oscillations means that most of the effect cancels out due to the integral contained in (\ref{eqn:effect-est}).

\begin{wrapfigure}{Rh}{0.6\textwidth}
\centering
\includegraphics[width=0.9\textwidth]{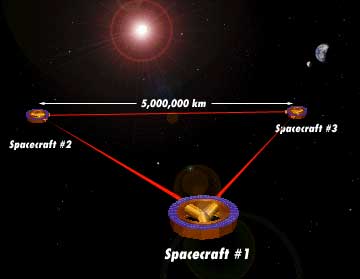}
\caption{The three LISA spacecraft showing a typical separation (credit: NASA).}
\label{fig:LISA}
\end{wrapfigure}

However, there is an alternative, which makes use of a proposed space-bourne detector called {\em LISA} (for Laser Interferometer Space Antenna), and considers the final merger event between two supermassive black holes.

Suppose we had two $10^8 \msun$ black holes merging at a distance from us of $500\mpc$. For a head-on collision, this would produce a `pulse' lasting about 1/2 hour. The relative velocity induced by this event in the two arms of the LISA probe (about 5 million ${\rm \, km}$ separation --- see Fig.~\ref{fig:LISA}), would be about 0.2 nanometer per second, and although sounding tiny, I believe is well within the velocity sensitivity of LISA (see e.g.\ \cite{danzmann2003lisa}). This is an exciting possibility, but of course depends on the final coalescence occurring whilst LISA was observing, and the odds on this would have to be established.

As a second example of an astrophysical possibility for obtaining `velocity memory', we could consider two stars  $1 {\rm \, pc}$ apart at $70 {\rm \, pc}$ from the same merger event as just discussed (i.e.\ deep inside the central region of the merging galaxies). The velocity kick in this case would be about $10 \kms$, which sounds eminently observable, but presumably is unfortunately completely overwhelmed by everything else going on nearby.

\section{Discussion, and possible theoretical relevance of `velocity memory' effect}

Since the focus of this article is on Geometric Algebra and Gauge Theory Gravity as applied to gravitational waves, we will shortly discuss in more detail how gravitational waves appear in GTG and the calculations which lead to the effects on particles discussed so far. Before that it may be useful to give a short sketch of what other people have said on the topic of `velocity memory', and also on its possible theoretical relevance.

It is not altogether clear whether one can say that people have realised before that particles initially at rest would be given a `kick' by a passing gravitational wave, and acquire non-zero velocities. Because in the standard $TT$ gauge the effect is entirely absent at linear level, most astrophysicists will probably not have thought in these terms. Staying in a $TT$-like gauge, but working exactly, then as we have seen, the effect is disguised as a progressive proper distance change, leading possibly to a collapse to a singularity after the particle has passed. This aspect is certainly discussed in e.g.\ the book by Misner, Thorne \& Wheeler, \cite{1973grav.book.....M}, Chap.~35. They highlight there how a better coordinate system was introduced in 1962 by Ehlers \& Kundt \cite{ehlers1962exact}, and this is basically a version of the {\em Brinkmann gauge} \cite{brinkmann1925hw}, and also a precursor of what are called $pp$-waves (for plane polarized).

So is the effect interesting? It may have theoretical relevance as another possible example of {\em gravitational wave memory}. The idea of this is that after a wave passes through, it leaves a permanent change in the proper distance, hence strain, in the observer's neighbourhood.
(At least part of this is easy to understand in terms of the change in `$M$' due to loss of mass in the merger.)
It appears this effect was first discussed in 1974 by Zeldovich \& Polnarev \cite{1974SvA....18...17Z},
and was then treated in detail in 1992 by Thorne \cite{1992PhRvD..45..520T}. Furthermore a `velocity memory' version of this effect was discussed by Grishchuk and Polnarev in 1989 (\cite{Grishchuk:1989qa}), although this was carried out within a linearised approach and with specialised sources (such as  star passing through the accretion disk surrounding a black hole), rather than being a general effect arising from an exact approach, as here. It will be interesting in the future to compare our predictions with the special cases they were considering, so as to understand the relation between the two in more detail.

\begin{wrapfigure}{Rh}{0.525\textwidth}
\centering
\includegraphics[width=0.85\textwidth]{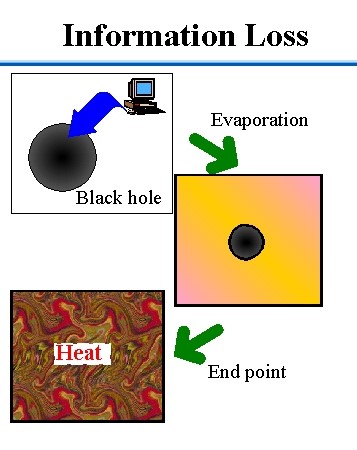}
\caption{Schematic illustration of the information loss problem for black holes ({\small credit: Gabor Kunstatter, University of Winnipeg}). \label{fig:bh-info-prob}}
\end{wrapfigure}
What is developing currently, is a very interesting possible theoretical connection between this memory effect, and the information loss problem for black holes.
This latter problem is the famous one concerning the eventual fate of the `information' corresponding to material which formed the black hole, schematically indicated in Fig.~\ref{fig:bh-info-prob}. Suppose that we have some object with a rich structure such as a computer which falls into a black hole, thereby increasing its mass. Eventually the black hole will disappear by emission of Hawking radiation, which in terms of its quantum field theory description is purely thermal, and only has random correlations, meaning that the information describing the complex objects from which the black hole was formed, appears to have been lost.

Putting it more mathematically, the process of black hole formation and evaporation appears to be {\em non-unitary} (information is lost), hence disagrees with the fundamental principles of quantum mechanics.

Without explicitly demonstrating how it might work, Strominger \& Zhiboedov \cite{Strominger:2014pwa} briefly discussed the link between the information paradox, and a result they had found which linked gravitational wave `memory' with the {\em Bondi-Metzner-Sachs Group}, which is the group of possible transformations of a spherical gravitational wave signal at infinity. They also showed that the GW displacement memory formula is equivalent to a formula for soft graviton production first given by Weinberg in 1965 \cite{1965PhRv..140..516W}.

Then Hawking, Perry \& Strominger \cite{Hawking:2016msc} linked this with what they called black hole `soft hair', and the storage of information about the formation of the black hole holographically at infinity. The question then arising, is whether our `velocity' rather than `displacement' memory effect is another channel by which information about what went into the black hole might also be stored (effectively at infinity) as a result of the waves emitted at black hole formation. This, and other types of memory effects, such as the `spin memory' recently put forward by Pasterski, Strominger \& Zhiboedov \cite{2016JHEP...12..053P}, are presumably going to be important in this `information budget', but the precise way in which this happens is so far unclear.
Returning to the question as to whether the {\em velocity} memory effect has been clearly identified before, this can certainly be answered in the context of the last two years, since shortly after I first spoke about this effect in a couple of meetings in April/May 2017, I discovered that Gary Gibbons, Peter Horvathy and co-workers had independently been looking at this, and their first full paper on this appeared later in 2017 \cite{2017PhLB..772..743Z}. This clearly identifies the effect discussed here, and in the eventual final version of their paper is expressed unambiguously in the Brinkmann gauge. Thus even though Geometric Algebra and Gauge Theory Gravity are evidently not necessary for discovering and understanding this effect, and finding out the best coordinate system in which to work with it (and more generally with plane fronted gravitational waves), it is nevertheless noteworthy that the GA/GTG approach was able to reach so quickly the physical answers which had taken about 50 years of the development of the subject to be teased out conventionally!

\section{Detailed gravitational wave discussion}

As we have just seen, Gauge Theory Gravity (GTG), which uses Geometric Algebra for its mathematical expression, can give simple and compact forms of solution for gravitational waves, both in exact and linearised form, and can aid in physical understanding of the waves. This occurs in a similar fashion to the way in which electromagnetic waves are simplified using Geometric Algebra (GA), and indeed the similarities with electromagnetism are emphasised in this approach. Separately, use of the Brinkmann metric in General Relativity (mentioned in the Introduction and the last section), rather than the standard Rosen or Bondi forms, has long been understood to have advantages for exact gravitational plane waves. However, for standard astrophysical and cosmological approaches to linearised waves, the Brinkmann gauge does not figure, certainly within the literature familiar to those working at the applied rather than mathematical end of gravitational wave investigations.

What we wish to do here, in the final parts of this contribution, is to present the mathematical underpinnings of the results presented in outline form above, and to relate the GTG approach clearly to the General Relativity approach, so that practitioners from either side can understand the new features being discussed and where the solutions come from.

\subsection{Exact waves in Minkowski space}
\label{eqn:sect-exact-waves}

For this, as described above, we can use a very simple form of $h$-function, basically a $(t,z)$ version of the $(t,r)$ Advanced Eddington Finklestein metric. 
We use
\be
\hob(a)=a-\half H a\dt e_+ e_+
\label{eqn:h-function}
\ee
where $e_+=e_t+e_z$, and $H$ is, initially at least, a general function of spacetime position. Using the results in \cite{2004gr.qc.....4081D}, leads to the following $\Omega(a)$ function
\be
\Omega(a)=\omega(a)=-\half \left(a\dt e_+\right) \grad H\wdg e_+
\ee

The traction of $\Omega(a)$ is
\be
\da \Omega(a)=-\half e_+ \dt \grad H e_+ =-\textstyle{\frac{1}{4}} e_+\left(\grad H\right) e_+
\ee
which is therefore the reflection of $\grad H$ in the null vector $e_+$, or equivalently the projection of $\grad H$ onto $e_+$. The latter means that if $H$ is of the separated wavelike form $H(t,x,y,z)=G(t-z) f(x,y)$, which we henceforth assume, then $\da \Omega(a)=0$. This leads to the following very simple form of the Einstein tensor (see Section~\ref{sect:back-curv-and-em} for definition)
\be
\begin{aligned}
\clg(a)&=-\half \left(e_+ \dt a\right) e_+ G(t-z) \grad^2 f\\
&=-\textstyle{\frac{1}{4}}e_+ a e_+ G(t-z) \grad^2 f
\end{aligned}
\ee
The Einstein equations are therefore solved if the 2d Laplacian $\grad^2 f(x,y)$ vanishes. We will want to do this in such a way as to leave a non-vanishing Weyl, so that there is genuine spacetime curvature, but at the same time the Weyl should be constant over the whole $(x,y)$ plane, since we want a plane-fronted wave.

With the ansatz $H(t,x,y,z)=G(t-z) f(x,y)$, and using the GTG definition of the Weyl
\be
\clw(a\wdg b)= \clr(a\wdg b)-\half\left(\clr(a)\wdg b + a\wdg\clr(b)\right)+{\textstyle \frac{1}{6}}a\wdg b \, \clr
\ee
we find the following compact form:
\be
\clw(B) = -\textstyle{\frac{1}{8}} e_+\grad\left(B\grad H\right)e_+
\ee
At this point it is convenient to switch to polar coordinates (equivalently cylindrical) in the $(x,y)$ plane, using $x=\rho\cos\phi$ and $y=\rho\sin\phi$. Making this substitution, employing $\grad^2 f$=0, and asking that the Weyl be constant over the $(x,y)$ plane, leads to the following two independent solutions for $f(x,y)$:
\be
f(x,y)=\rho^2\cos2\phi \quad \text{and} \quad f(x,y)=\rho^2\sin2\phi
\ee
It is convenient to borrow some of the freedom from the $G(t-z)$ part of $H(t,x,y,z)=G(t-z) f(x,y)$, in order to combine these solutions in the form
\be
H(t,x,y,z)=G(t-z)\rho^2\cos\left(2\left(\phi-\phi_0\right)\right)
\label{eqn:H-sol}
\ee
where $\phi_0$ is an arbitrary function of $t-z$. This angle represents the {\em polarization direction} of the wave. Specifically, we can define the polarization direction in the $(x,y)$ plane as the unit vector
\be
e_{\perp}=\cos(\phi_0(t-z))e_x+\sin(\phi_0(t-z))e_y
\ee
and then we find for the Weyl corresponding to the solution (\ref{eqn:H-sol})
\be
\clw(B)=-\half G(t-z) \, e_+ e_{\perp} B e_{\perp} e_+
\label{eqn:Weyl-compact}
\ee
As said above, this is very neat in showing us how the input bivector $B$ is reflected in the bivector $e_+ e_{\perp}$, which encodes both the direction of propagation in spacetime, and the direction of polarization. Equivalently, we can think of $B$ as being reflected in the direction representing polarization, and then projected down the null vector $e_+$. We can tie in what we have just found with the expressions for the Weyl tensor for gravitational waves given in \cite{1998RSPTA.356..487L}, via
\be
\clw(B) = -\textstyle{\frac{1}{4}} G(t-z)\left\{ \cos(2\phi_0) \clw^+(B)+ \sin(2\phi_0) \clw^{\times}(B)\right\}
\label{eqn:Weyl-explicit}
\ee
where
\be
\begin{aligned}
\clw^+(B) &= e_+\left(e_x B e_x -e_y B e_y\right) e_+\\
\text{and} \quad
\clw^{\times}(B) &= e_+\left(e_x B e_y +e_y B e_x\right) e_+
\end{aligned}
\ee

The way that the polarization angle is given by $\phi_0$, whereas the solution for $H$, equation (\ref{eqn:H-sol}), and the components of the Weyl, equation (\ref{eqn:Weyl-explicit}), rotate through $2\phi_0$, is of course a consequence of the `spin-2' nature of gravitational radiation, and it is interesting to see it arising here due to the fact we are {\em reflecting} in the polarization direction.

We can get further insight into these rotation properties by considering the position and rotation gauge properties of the Weyl tensor in the form (\ref{eqn:Weyl-compact}), and of the $h$-function (\ref{eqn:h-function}). We can do this for a rotation in the $(x,y)$ plane through angle $\phi_0$, where we treat this as a constant. This is allowed, since as we have seen, allowing it to be an arbitrary function of $t-z$ amounts to linearly superposing solutions in which the amplitude part $G(t-z)$ has been redefined to accommodate this variation. That we can add solutions linearly follows from the fact that $H$ only appears linearly in all the equations, a distinctive feature of this approach.

We thus define the following rotor:
\be
R= \exp(-\half  \phi_0 I \sigma_3)
\ee
which rotates $e_x$ to $e_{\perp}$.

The overall gauge change we are going to carry out is the one described near equation (6.11) in \cite{1998RSPTA.356..487L}, namely rotating $\hob(a)$ to $R\hob(a)\Rrev$, and then carrying out the position gauge transformation to the backrotated position
\be
x'=\Rrev x R
\label{eqn:x-disp}
\ee
This would yield no overall change if $\hob(a)$ were cylindrically symmetric, but gives a new configuration otherwise. The $\underline{f}(a)$ function corresponding to the displacement $f(x)=x'$ is
\be
\underline{f}(a)=a\dt\grad\left(\Rrev x R\right)=\Rrev a R
\ee
if $\phi_0$ is constant. The rotated and displaced $\hob(a)$ function is thus
\be
\hob_{x'}\fob^{-1}(a)=R\hob\left(\Rrev a R,x'\right)\Rrev
\ee
for a general $\hob$, and for the specific form in (\ref{eqn:h-function}), then since $R e_+ \Rrev=e_+$, we get a new $\hob$ function of
\be
\hob'_x(a)=a-\half H(x') a\dt e_+ e_+
\ee
i.e.\ the only change is in the point of evaluation of $H$, which is rotated through $-\phi_0$ in the $(x,y)$ plane, in agreement with the result in equation (\ref{eqn:H-sol}). This will lead to a Weyl tensor of the form (\ref{eqn:Weyl-compact}) assuming the initial $H$ was for $\phi_0=0$, i.e.\ polarization direction along the $e_x$ axis.

For the Weyl tensor, as against the $\hob$-function, effectively the opposite happens. As a covariant object, we know it transforms under position and rotation gauge changes as discussed in Section 4 of \cite{1998RSPTA.356..487L}, i.e.
\be
\begin{aligned}
\text{Translations}&: \quad \clw'(B;x)=\clw(B;x') \\
\text{Rotations}&: \quad \clw'(B)=R\clw(\Rrev B R) \Rrev
\end{aligned}
\ee
In the current case, since $\clw$ has no position dependence in the $(x,y)$ plane, then it sees only the rotor gauge change, and assuming as for $\hob$ that $e_{\perp}$ starts as $e_x$, we obtain
\be
\begin{aligned}
\clw'(B) &= -\half G(t-z) R e_+ e_x \Rrev B R e_x e_+ \Rrev\\
&=-\half G(t-z) \, e_+ e_{\perp} B e_{\perp} e_+
\end{aligned}
\ee
as already found in (\ref{eqn:Weyl-compact}). Thus although the $\hob$ function does not see the rotor change, only the displacement to the backrotated position, while the Weyl tensor does not see the displacement, only the forward rotor transformation, the resulting Weyl tensors of course agree, showing that everything is set up consistently.

In a GR, as against GTG, context, of course all that would be visible is the change in the metric caused by the displacement gauge change --- all the rotor gauge changes would be invisible.

\subsection{Effects on particle motion}

We may write the GTG geodesic equation as
\be
\dot{v}+\omega(v)\dt v=\dot{v}+\Omega(\dot{x})\dt v=0
\ee
where $\dot{\ }$ is differentiation w.r.t.\ proper time along the path, and we have
\be
v=\hub^{-1}(\dot{x})
\ee
For a particle instantaneously at rest in the laboratory frame, therefore, and for small $H$, so that $v \approx \gamma_0$ as well, we can think of $-\omega(\gamma_0)\dt\gamma_0$ as the instantaneous force on the particle. For the current $\hob$-function and choice of $H$ given in (\ref{eqn:H-sol}), this evaluates to
\be
-G(\eta)\rho\cos\left(\phi-2\phi_0\right)\gamma_1+G(\eta)\rho\sin\left(\phi-2\phi_0\right)\gamma_2
\ee

\begin{figure}
\begin{center}
\includegraphics[width=0.45\textwidth]{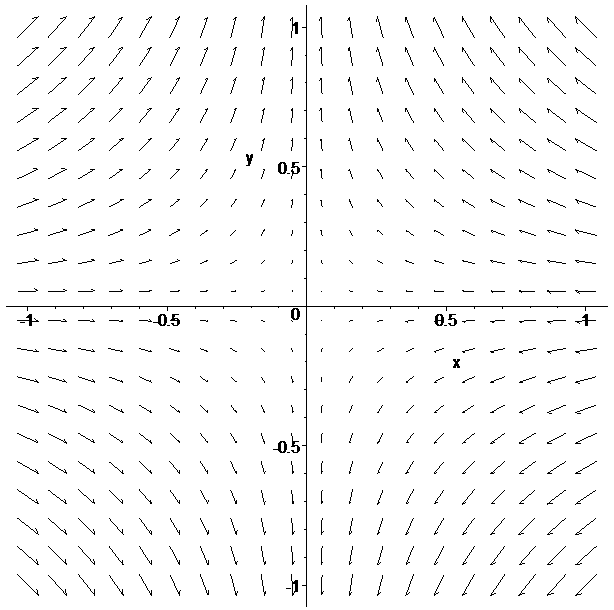}
\includegraphics[width=0.45\textwidth]{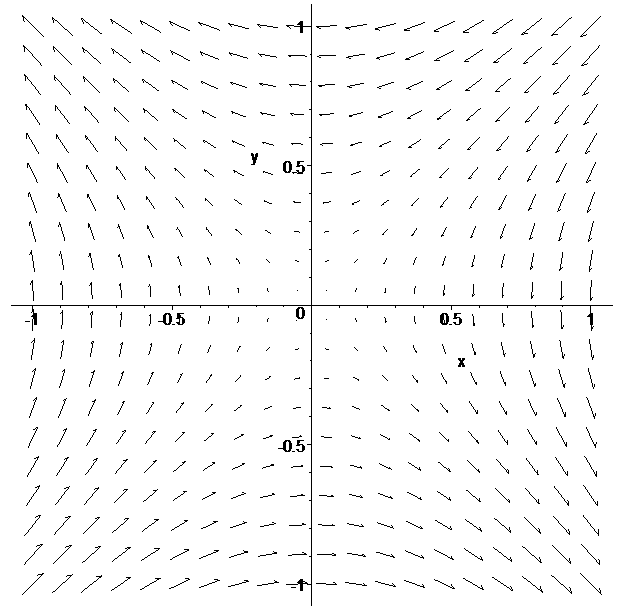}
\caption{Force vectors (normalised by $1/G(\eta)$) in the $(x,y)$ plane, for polarization angle $\phi_0=0$ (left) and $\phi_0=\pi/4$ (right).}
\label{fig:force-plots}
\end{center}
\end{figure}

We see that this force lies wholly in the $(x,y)$ plane, and shows us that the particles are going to be `jiggled' in this plane as the gravitational wave passes through. This gives a nice intuitive picture of the action of the wave on particles. We show in Fig.~\ref{fig:force-plots}
the force vectors in the plane (per unit $G(\eta)$) for the values $\phi_0=0$ and $\phi_0=\pi/4$.

It will clearly be worried about what happens for large values of $x$ and $y$, and whether our picture breaks down there. Also, a preferred centre seems to have been picked out where the force is zero (on the $z$ axis), and this needs understanding in the light of the desired planar rather than cylindrical nature of the wave, and we discuss these two crucial issues further below, once we have full solutions in hand for the particle motion.

To get the full solutions, it proves easiest in this case to work with the (traditional) second order equations in $\ddot{x}$, rather than finding the coupled first order equations for $v$ and $\dot{x}$. The results we find, where we write the coordinates $(t,x,y,z)$ as $(\sft,\sfx,\sfy,\sfz)$ to avoid any confusion with the 4d position $x$, and with $\eta$ still denoting $t-z=\sft-\sfz$, are
\be
\begin{gathered}
\ddot{\eta}=0, \quad \ddot{\sfx}=-G \dot{\eta}^2\left(\sfx \cos2\phi_0+\sfy \sin2\phi_0\right)\\
\ddot{\sfy}=G \dot{\eta}^2\left(\sfy \cos2\phi_0-\sfx \sin2\phi_0\right)\\
\ddot{z}=-\half\dot{\eta}\left(\left(\sfx^2-\sfy^2\right)G'\dot{\eta}+4G\left(\sfx\dot{\sfx}-\sfy\dot{\sfy}\right)\right)\cos 2\phi_0\\
-\half\dot{\eta}\left(2\sfx\sfy G'\dot{\eta}+4G\left(\sfx\dot{\sfy}+\sfy\dot{\sfx}\right)\right)\sin 2\phi_0
\end{gathered}
\label{eqn:geo-ddots}
\ee
It is evident from this, and from the symmetries of the force plots shown in Fig.~\ref{fig:force-plots}, that as regards particle motions, we can work w.l.o.g.\ in a system where the polarization angle $\phi_0=0$. This then simplifies these results to
\be
\begin{gathered}
\ddot{\eta}=0, \quad \ddot{\sfx}=-\sfx G \dot{\eta}^2, \quad \ddot{\sfy}=\sfy G \dot{\eta}^2\\
\ddot{z}=-\half\dot{\eta}\left(\left(\sfx^2-\sfy^2\right)G'\dot{\eta}+4G\left(\sfx\dot{\sfx}-\sfy\dot{\sfy}\right)\right)
\end{gathered}
\label{eqn:second-derivs}
\ee
The very nice things about these equations, is that the first implies $\dot{\eta}$ is constant, and then the $x$ and $y$ equations may be solved separately from everything else, to give the motion in the $(x,y)$ plane as driven by the wave amplitude $G(\eta)$. We give a concrete example of this shortly.

As a check of these equations, we can ask if they satisfy the constraint that the length of the velocity 4-vector is maintained as 1. We have (for $\phi_0=0$)
\be
\begin{gathered}
v=\left\{\left(1+\half\left(\sfx^2-\sfy^2\right)G\right)\dot{\eta}+\dot{\sfz}\right\}\gamma_0+\dot{\sfx}
\gamma_1+\dot{\sfy}\gamma_2\\
\left\{\half\left(\sfx^2-\sfy^2\right)G\dot{\eta}+\dot{\sfz}\right\}\gamma_3
\end{gathered}
\ee
The squared length of this is
\be
v^2=2\dot{\eta}\dot{\sfz}-\dot{\sfx}^2-\dot{\sfy}^2+\left(1+\left(\sfx^2-\sfy^2\right)G\right)\dot{\eta}^2=1
\label{eqn:v-constraint}
\ee
so this forms a constraint that for example can be used to find $\dot{\eta}$ at an initial time, given the other velocities then. Differentiating the constraint and substituting in the second derivatives (\ref{eqn:second-derivs}), we get zero, showing that indeed the constraint is compatible with the derivatives.

Now suppose that we have managed to solve the $\sfx$ and $\sfy$ equations in (\ref{eqn:second-derivs}) for some given driving function $G(\eta)$. We would then like to find the motion in $\sfz$ as well, and in principle this could be done via integration of the first order equation (\ref{eqn:v-constraint}). Perhaps surprisingly, there is in fact a general expression for $\sfz$ available which automatically satisfies (\ref{eqn:v-constraint}), and which we can use immediately to find $\sfz$ from the $\sfx$ and $\sfy$ motions. If we write the affine parameter along the path with respect to which we are differentiating as $s$, the result can be written as
\be
\sfz=\frac{1}{4a}\left(\frac{d}{ds}\left(\sfx^2+\sfy^2\right)-2\left(a^2-1\right)s\right) + {\rm const.}
\label{eqn:z-from-xy}
\ee
where $a=\dot{\eta}=d\eta/ds$ is a constant. Remarkably the relation (\ref{eqn:z-from-xy}) does not explicitly depend on the gravitational wave amplitude $G(\eta)$, but holds generally. In particular it automatically satisfies the constraint (\ref{eqn:v-constraint}) and the $\ddot{\sfz}$ relation in (\ref{eqn:second-derivs}), by virtue of the (decoupled) $\ddot{\sfx}$ and $\ddot{\sfy}$ relations in (\ref{eqn:second-derivs}). Thus part of the $\sfz$ motion is driven by the rate of change of cylindrical radial distance, $\sqrt{\sfx^2+\sfy^2}$, and the rest is a constant linear motion driven by $a^2-1=\dot{\eta}^2-1$. We can complete the solution by noting that since $\eta=\sft-\sfz=as$ we can derive $\sft$ simply by adding $as$ to the expression for $\sfz$ in (\ref{eqn:z-from-xy}).

Finally, as regards general properties of the motion, one can show that, despite an explicit origin, at which there is zero force, the motion is in fact homogeneous over the entire $(\sfx,\sfy)$ plane.

\subsection{Analytic solution in the harmonic case}
\label{sect:harm-sol}

Having established some general properties of the geodesics, we now look at solutions in some specific cases, to illustrate new features which arise, starting with the case of a {\em harmonic} wave.

So here the gravitational wave has the driving function
\be
G(\eta)=A \cos(\omega \eta) = A \cos( a s\omega)
\ee
for constant amplitude $A$ and angular frequency $\omega$.
One finds that the solutions for the $\sfx$ and $\sfy$ particle motion equations for a particle starting at rest at position $(\sfx_0,\sfy_0)$ in the $(\sfx,\sfy)$ plane are the {\em Mathieu functions}
\be
\sfx=\sfx_0 {\rm \, MathieuC}\left(0,-\frac{2A}{\omega^2},\frac{\omega \eta}{2}\right), \quad \sfy=\sfy_0 {\rm \, MathieuC}\left(0,\frac{2A}{\omega^2},\frac{\omega \eta}{2}\right)
\ee
where we are using the Maple notation for these functions. In particular ${\rm MathieuC}(b,q,v)$ is one of the two linearly independent solutions of the Mathieu equation
\be
u''+\left(b-2q\cos(2v)\right)u=0
\ee
for a function $u(v)$, and is specified by having zero derivative at $v=0$ and value 1 there.

If one plots these out, this leads to a surprise.
\begin{figure}[h]
\centering
\setlength{\lineskip}{\medskipamount}
\subcaptionbox{Initial motion for a particle starting at rest at $(0.8,0.2)$ in the $(\sfx,\sfy)$ plane. The gravitational wave angular frequency and amplitude are $\omega=5.0$ and $A=0.1$ respectively. \label{fig:harm-motiona}}{\includegraphics[width=0.59\textwidth]{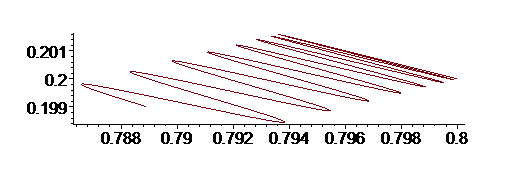}}\hfill
\subcaptionbox{Long time development of the motion. Here the $\sfx$ coordinate is plotted versus affine path parameter $s$.\label{fig:harm-motionb}}{\includegraphics[width=0.39\textwidth]{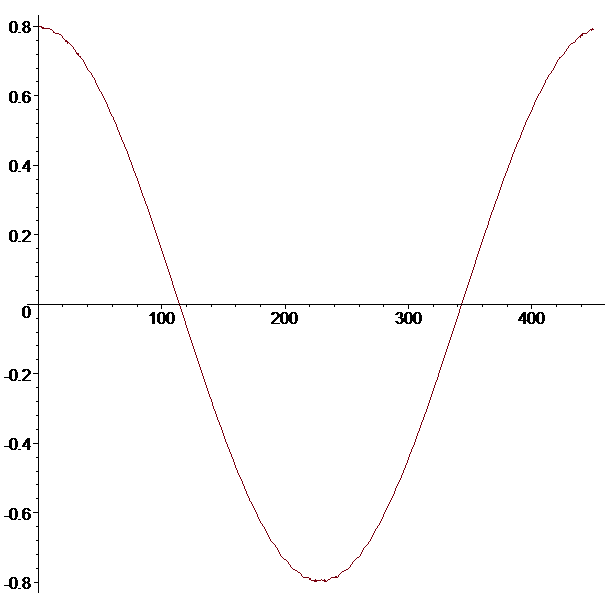}}
\caption{Particle motion in a harmonic wave.} \label{fig:harm-motion}
\end{figure}
The particle initially oscillates back and forth along a constant line, (top right portion of Fig.~\ref{fig:harm-motiona}) but then starts gradually to move off --- eventually approaching the origin, passing through it to the other side and then returning to the original position, at which point the long term cycle repeats over again (Fig.~\ref{fig:harm-motionb}). Note the small oscillations visible near the peaks and troughs in the plot of long-term $\sfx$ position are real, and correspond to positions where the continuous (and expected) `jiggling' at the wave frequency are visible.

The long-term motion, and in particular the fact it cycles through the origin, are the manifestation of what in previous approaches would have been thought of as coordinate singularities, corresponding to collapse of the coordinate system (in a particular direction) to a point, rather than as real long term excursions by the particle.

\subsection{Impulsive waves}

\label{sect:part-motion}

Continuous oscillation is of interest in terms of real gravitational waves, but it is also interesting to consider a non-oscillating `bump' or `burst'. This might come about, for example, due to the nearly head-on collision of two black holes (see e.g.\ \cite{PhysRevD.71.124042} for the fully head-on case).

The essence of a bump or burst is that it should only last a finite time and is zero outside those times. Such a function is difficult to implement numerically whilst keeping the function and all derivatives continuous, so as a starting point we consider what happens for a {\em Gaussian} shaped profile of the form
\be
G(\eta)=A\exp\left(-\frac{\eta^2}{2\sigma^2}\right)=A\exp\left(-\frac{(as)^2}{2\sigma^2}\right)
\ee
where $s$ is the affine parameter, and $a=\dot{\eta}$ as above. This is effectively zero a few sigma outside the core.

In the example already shown in Figs.~\ref{fig:2dcaust} and \ref{fig:3dcaust}, we start the particles in a ring of radius $0.8$ in the $(\sfx,\sfy)$ plane, and have a Gaussian wave impinging on them from the negative $z$ direction with $A=0.1$ and $\sigma=2$. The motion is followed for a total affine parameter elapse of $s=8$. As already discussed, the motions are interesting, and clearly show the development of caustics.

These plots were produced by numerical integration, and for a function that is guaranteed smooth. The latter point means we do not have to doubt the results on the ground that maybe some impulse to the particle has inadvertently come about from a discontinuity in the derivative of the function representing the wave. The function is smooth, returns to exactly what its value was before the `burst' started (i.e.\ effectively zero), as do all its derivatives, and yet we still have a net impulse imparted to the particles.

With reassurance from this result in hand, we can move forward to a case which can be treated more easily in an analytic fashion. This is for a pulse of the form
\be
G(\eta) = \begin{cases} d^2 \, e^{-c\eta} & \eta>0 \\ d^2 \, e^{c\eta} & \eta<0 \end{cases}
\label{eqn:impluse-form}
\ee
where $d$ and $c$ are constants defining the amplitude and width of the pulse. The general solution of the geodesic equations for this case (taking $\eta>0$ temporarily) is
\be
\sfx = C_1 J_0\left(\frac{2d}{c}e^{-\frac{c\eta}{2}}\right) + C_2 Y_0\left(\frac{2d}{c}e^{-\frac{c\eta}{2}}\right)
\label{eqn:impluse-sol}
\ee
where $J_0$ and $Y_0$ are the zeroth order Bessel functions and $C_1$ and $C_2$ are constants. We can obtain the general solution for $\eta<0$ by switching the sign of $c$. There is a similar formula for the $\sfy$ behaviour, and we obtain the $\sfz$ behaviour from our general formula (\ref{eqn:z-from-xy}). We can see that interestingly the arguments of the Bessel functions in (\ref{eqn:impluse-sol}) are a type of `square root' of the applied impulse (\ref{eqn:impluse-form}) and this is happening due to $G$ being like a varying `$\omega^2$' in the underlying approximately harmonic equations we are working with.

We can now get a general expression for the velocity induced by the pulse by (a) setting up the particle motion to start from rest at some point $\eta=\eta_0$ a long way before the wave arrives (so $\eta_0<0$); (b) evaluating this solution and its first derivative at $\eta=0$ and (c) matching these to a general solution at $\eta>0$ to get the future behaviour of the particle. The result is
\begin{equation}
\begin{aligned}
\sfx(\eta)&=-\sfx_0\, \left( { Y_0} \left( 2\,d\tau \right) { J_0} \left( 2
\,d\tau\,{{\rm e}^{-1/2\,{\frac {\eta}{\tau}}}} \right) { Y_1}
 \left( -2\,d\tau\,{{\rm e}^{1/2\,{\frac {{\eta_0}}{\tau}}}}
 \right)
 { J_1} \left( 2\,d\tau \right)\right.\\
 &-{ Y_0} \left( 2\,d\tau
 \right) { J_1} \left( 2\,d\tau\,{{\rm e}^{1/2\,{\frac {{\eta_0}}{
\tau}}}} \right) { J_0} \left( 2\,d\tau\,{{\rm e}^{-1/2\,{\frac {
\eta}{\tau}}}} \right)
{ Y_1} \left( -2\,d\tau \right) \\
&-{ J_1}
 \left( 2\,d\tau \right) { J_1} \left( 2\,d\tau\,{{\rm e}^{1/2\,{
\frac {{\eta_0}}{\tau}}}} \right) { Y_0} \left( 2\,d\tau\,{{\rm e}
^{-1/2\,{\frac {\eta}{\tau}}}} \right) { Y_0} \left( -2\,d\tau
 \right) \\
 &-2\,{ J_1} \left( 2\,d\tau \right)
 { Y_0} \left( 2\,d\tau
\,{{\rm e}^{-1/2\,{\frac {\eta}{\tau}}}} \right) { Y_1} \left( -2\,d
\tau\,{{\rm e}^{1/2\,{\frac {{\eta_0}}{\tau}}}} \right) { J_0}
 \left( 2\,d\tau \right) \\
 &+{ J_1} \left( 2\,d\tau\,{{\rm e}^{1/2\,{
\frac {{\eta_0}}{\tau}}}} \right) { J_0} \left( 2\,d\tau\,{{\rm e}
^{-1/2\,{\frac {\eta}{\tau}}}} \right) { Y_1} \left( 2\,d\tau
 \right) { Y_0} \left( -2\,d\tau \right) \\
 &+{ J_1} \left( 2\,d\tau\,
{{\rm e}^{1/2\,{\frac {{\eta_0}}{\tau}}}} \right) { Y_0} \left( 2
\,d\tau\,{{\rm e}^{-1/2\,{\frac {\eta}{\tau}}}} \right) { J_0}
 \left( 2\,d\tau \right) { Y_1} \left( -2\,d\tau \right) \\
 &+\left({ J_0}
 \left( 2\,d\tau \right) { Y_1}
 \left( 2\,d\tau \right) { J_0}
 \left( 2\,d\tau\,{{\rm e}^{-1/2\,{\frac {\eta}{\tau}}}} \right) {
Y_1} \left( -2\,d\tau\,{{\rm e}^{1/2\,{\frac {{\eta_0}}{\tau}}}}
 \right)  \right)/\\ &
 \bigg(
 \left( { J_0} \left( 2\,d\tau\,{{\rm e}^{1/2\,{
\frac {{\eta_0}}{\tau}}}} \right)
{ Y_1} \left( -2\,d\tau\,{
{\rm e}^{1/2\,{\frac {{\eta_0}}{\tau}}}} \right) +{ Y_0} \left( -2
\,d\tau\,{{\rm e}^{1/2\,{\frac {{\eta_0}}{\tau}}}} \right) { J_1}
 \left( 2\,d\tau\,{{\rm e}^{1/2\,{\frac {{\eta_0}}{\tau}}}} \right)
 \right) \\
& \left( -{ J_0} \left( 2\,d\tau \right) { Y_1}
 \left( 2\,d\tau \right) +{ Y_0} \left( 2\,d\tau \right) { J_1}
 \left( 2\,d\tau \right)  \right)
 \bigg)
 \end{aligned}
 \label{eqn:imp-exact}
\end{equation}

In this equation we have replaced the rate constant $c$ in the expression for the impulse, (\ref{eqn:impluse-form}), with a time constant $\tau=1/c$. This expression looks complicated, but the `live' dependencies are just the terms
${ J_0} \left( 2\,d\tau\,{{\rm e}
^{-1/2\,{\frac {\eta}{\tau}}}}\right)$ and ${ Y_0} \left( 2\,d\tau\,{{\rm e}
^{-1/2\,{\frac {\eta}{\tau}}}}\right)$, the rest being just fixed constants.
We have given this expression in full, since we want to draw attention to how we can use it to calculate the velocity deflection of a particle in the impulsive limit.

\begin{figure}
\centering
\setlength{\lineskip}{\medskipamount}
\subcaptionbox{Overview of motion. \label{fig:imp-motiona}}{\includegraphics[width=0.49\textwidth]{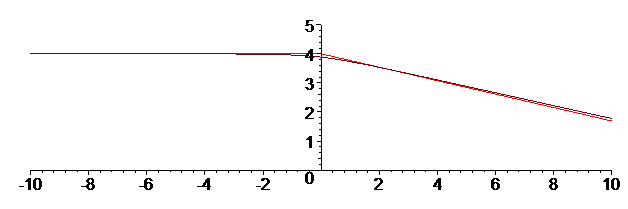}}\hfill
\subcaptionbox{Detail near $\eta=0$, when the wave arrives.\label{fig:imp-motionb}}{\includegraphics[width=0.49\textwidth]{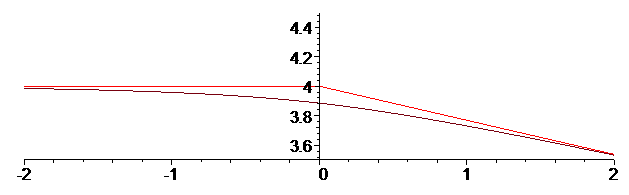}}
\caption{$\sfx$ motion of a particle starting at rest at $\sfx=4$ in an impulsive gravitational wave with strength $d=0.17$ and time constant $\tau=1$. The motion is followed from $\eta=-10$ to +10. Black shows the exact expression (\ref{eqn:imp-exact}) and red shows the approximation from (\ref{eqn:imp-approx}). } \label{fig:imp-motion}
\end{figure}

First of all, we use this expression to get the exact motion of a particle initially at rest being acted on by an impulsive wave in an illustrative case shown in Fig.~\ref{fig:imp-motion}.
This exact result is the black curve in the two plots. Seeing that the change in velocity is so nearly impulsive we now seek an approximation to (\ref{eqn:imp-exact}) which can capture the dependence of this velocity change on the parameters of the problem. We see that the important aspect comes from the behaviour of ${ J_0} \left( 2\,d\tau\,{{\rm e}
^{-1/2\,{\frac {\eta}{\tau}}}}\right)$ and ${ Y_0} \left( 2\,d\tau\,{{\rm e}
^{-1/2\,{\frac {\eta}{\tau}}}}\right)$ at large $\eta$ (as compared to $\tau$). We know that for small $\theta$
\be
J_0(\theta) \approx 1-\frac{\theta^2}{4}
\ee
while
\be
\quad Y_0(\theta)\approx 2\,{\frac {-\ln  \left( 2 \right) +\ln  \left( \theta \right) +\gamma}{\pi
}}-1/2\,{\frac { \left( -\ln  \left( 2 \right) +\ln  \left( \theta \right)
-1+\gamma \right) {\theta}^{2}}{\pi }}
\ee
where $\gamma$ is the Euler-Mascheroni constant ($\approx 0.5772$). As a first step we thus substitute $\theta=2\,d\tau\,{{\rm e}
^{-1/2\,{\frac {\eta}{\tau}}}}$ into these expressions, and then use them to substitute for ${ J_0} \left( 2\,d\tau\,{{\rm e}
^{-1/2\,{\frac {\eta}{\tau}}}}\right)$ and ${ Y_0} \left( 2\,d\tau\,{{\rm e}
^{-1/2\,{\frac {\eta}{\tau}}}}\right)$ in (\ref{eqn:imp-exact}). Perhaps surprisingly, the term that emerges as the dominant one at large positive $\eta$, is due to the $\ln(\theta)$ in the first term of the small $\theta$ approximation for $Y_0$!

With this in hand, the further approximation we need comes from noting that there is a dimensionless quantity which measures the `effective strength' of the pulse. From equation (\ref{eqn:H-sol}) which relates $H$ to $G$, and from (\ref{eqn:h-function}), which gives us the form of metric, we can see that $H$ is dimensionless, and that therefore $G$ has units of $L^{-2}$. (Here we are using natural units, in which everything can be expressed in terms of a single length scale.) The $d^2$ in the definition of the pulse profile, equation (\ref{eqn:impluse-form}), thus has dimensions $L^{-2}$ also, so when $d$ is combined with the time constant $\tau$ to form $d\tau$ the combination is in fact dimensionless. (This makes sense since the arguments to the Bessel functions should be dimensionless.) We can now set $\epsilon=d\tau$ and carry out an expansion of the quantity we found from (\ref{eqn:imp-exact}) using the above $J_0$ and $Y_0$ approximations, in powers of $\epsilon$. This yields the answer
\be
\sfx \approx \sfx_0\left(1-2\tau \eta d^2\right)
\label{eqn:imp-approx}
\ee
valid for $\eta\gg \tau$, where $\sfx_0$ is the original $\sfx$ position of the particle. Now carrying out the same process for $\sft=\sfz+\eta$, where $\sfz$ is given by the analytic expression (\ref{eqn:z-from-xy}), we find that there is no perturbation to $\sft$ at the same order, hence the change in velocity of the particle is simply
\be
\Delta v = -2 \sfx_0 \tau d^2
\ee
This ties in with our general result (\ref{eqn:effect-est}) since
\be
 - \sfx_0 \int G \, d\eta =-2 \sfx_0 \int_0^\infty \exp\left(-\frac{\eta}{\tau}\right) d\eta \, d^2= -2 \sfx_0 \tau d^2
\ee
for this case.

The next example it is worth to look at is a combination of the last two, which is for an oscillating signal bounded in time by an envelope which drops smoothly to 0 for large $|\eta|$. We can call this a `harmonic pulse'. Using this type of $G(\eta)$ it is possible to verify that the integral in the general formula (\ref{eqn:effect-est}) does the correct thing, i.e.\ that the deflection and velocity are proportional to the {\em integral} of $G$ over the pulse, and not (e.g.) to its value at the centre.

\subsection{`Background curvature' and electromagnetic waves}
\label{sect:back-curv-and-em}

We now want to discuss in a more general context, the `square root' solution for particle motion as given in (\ref{eqn:impluse-sol}), and also via this to examine some comments relating to exact gravitational waves made in the famous book by Misner, Thorne \& Wheeler (MTW), \cite{1973grav.book.....M}, with which we appear to disagree.

In Sections 35.9 to 35.12 of MTW, they discuss their viewpoint on exact gravitational waves, and in particular suggest that the energy density of the `ripples' causes a long term background curvature. This is argued particularly in relation to the same effect which they say happens for pure electromagnetic waves. To investigate this for both electromagnetic waves and gravitational waves jointly, we now suppose that we have both EM and gravitational plane waves moving in the same ($z$) direction, and we will start by using the same gauge as MTW use. This is an exact version of the TT gauge, of one of the forms discussed above, but which we have not so far explicitly specified. We write it first in metric form as
\be
ds^2=dt^2-L^2\left(e^{2\beta} dx^2 + e^{-2\beta} dy^2 \right) -dz^2
\ee
Here $L$ and $\beta$ are functions of $\eta=t-z$ (in MTW, our $\eta$ is called $u$, but otherwise the notation here is the same as used in MTW).

To include electromagnetic waves in the analysis, we need to understand how electromagnetism (EM) appears within the GA and GTG approach. This is described in Section~7 of \cite{1998RSPTA.356..487L}, but a brief summary, sufficient for our purposes here, is that the covariant version of the usual Faraday tensor is a bivector $\clf$, which satisfies the `curved space' Maxwell equations
\be
\cld \clf = \clj
\ee
where $\clj$ is the covariant current vector. This is really quite simple! The contribution to the total stress energy tensor (SET) given by the EM field is
\be
\tau_{\rm EM}(a) = -\half \clf a \clf
\ee
which is again quite simple. Our discussion of the gravitational field equations above was for the vacuum case. In the case where the matter and fields give rise to a non-zero SET, then the $\partial_{\hob(a)}$ equation becomes
\be
\clg(a) \equiv \clr(a) - \half a \clr = 8 \pi \tau_{\rm tot}(a)
\label{eqn:G-def}
\ee
where the {\em Ricci tensor} $\clr(a)$ is defined by
\be
\clr(a)=\db \dt \clr(b \wdg a).
\ee
Here $\tau_{\rm tot}(a)$ is the total matter SET and the first equality in (\ref{eqn:G-def}) defines the {\em Einstein tensor} $\clg(a)$ as the `trace-reverse' of the Ricci tensor. (This follows since $\clr=\da\dt\clr(a)$ and $\da a = 4$.)

As an ansatz for the covariant Faraday bivector we use exactly the same form as we would use for a plane wave moving in the $z$-direction without gravitation (see e.g.\ Section~7.4 of \cite{d2003geometric}), namely
\be
\clf=F(\eta)\sigma_1\left(1-\sigma_3\right)
\label{eqn:wave-calf}
\ee
Here $F(\eta)$ is meant to be any scalar function of $t-z$, and gives the EM pulse or wave shape. (Note this is the same symbol as the `straight $F$' version of the Faraday bivector used in \cite{d2003geometric} and \cite{1998RSPTA.356..487L}, but since we will not be using the latter here, there is hopefully no possibility of confusion.)

It turns out that $F$ being a function just of $\eta$ means that the `curved space' Maxwell equations are automatically satisfied, and the master equation in this case is the single SET equation
\be
L'' + L \left( 4 \pi F^2 +{\beta'}^2 \right)=0
\label{eqn:Ldd-wave}
\ee
where a dash represents a derivative w.r.t.\ $\eta$.

It is perhaps remarkable that the potentially complicated scenario of an exact self-consistent combination of a gravitational wave and an electromagnetic wave travelling in the same direction, can be represented by such a simple equation! We now use it to investigate what MTW say about the separate EM and GW cases.
As part of this, however, we also need formulae for the Riemann entries in this case, since these provide the gauge-invariant information for the gravitational sector. We will give a compact expression for the full Riemann below, but here we can state that its magnitude is controlled by the expression
\be
\beta''+{\beta'}^2+\frac{L''}{L}+\frac{2L'\beta'}{L}
\ee

So suppose we have an EM field, and turn off the $\beta$ part of the metric (which means we have turned off the gravitational waves). The magnitude of the Riemann is then determined entirely by $|L''/L|$, and this is just $4\pi F^2$. This is therefore the magnitude of the curvature in this case, and is given entirely by whatever function of $\eta$ the EM plane wave is --- there is no evidence for a `background curvature' responding to the overall energy density of the waves --- we just have an immediate response dictated locally by the value of the EM wave itself at that point.

Now suppose we turn $\beta$ back on.
The logic employed at this part of MTW, would now indicate that we should take $\beta(\eta)$ as the physically important quantity representing the wave, and then the `background curvature' $L(\eta)$ will respond to this, in the same way as it did for the EM wave, via equation (\ref{eqn:Ldd-wave}). There is nothing wrong with equation (\ref{eqn:Ldd-wave}), but what we are asserting instead is that the physically meaningful quantity for the gravitational wave, is not the function $\beta$, but instead the entries in the Weyl or Riemann tensor. It is only the latter which are gauge-invariant, and therefore for a wave transporting information, it is these that we must concentrate on, not the gauge-dependent metric function $\beta$. Employing the SET equation, it turns out the Weyl entries are controlled by the function
\be
\beta''+\frac{2}{L}L'\beta'
\ee
and it is this that should be taken as corresponding to the actual physical waveform. Trying to set this equal to a harmonic function, for example, leads to some rather messy non-linear equations, but fortunately, there is a much simpler route available. Instead of working with the metric components used by MTW, let us instead use the following $h$-function entries:
\be
g_x=m(\eta) e_x, \quad g_y=n(\eta) e_y
\ee
Then it is easy to see that (at least for $m$ and $n$ both positive), the link with the $L$ and $\beta$ used by MTW is
\be
L=\sqrt{m n}, \quad \beta=\half \ln\left(\frac{m}{n}\right)
\ee
This substitution `linearizes' the equations and gives a neat split into the following two master equations, which deal simultaneously with the EM and GW cases:
\be
m'' + m \left(4\pi F^2+G\right)=0, \quad  n'' + n \left(4\pi F^2-G\right)=0
\label{eqn:mn-master}
\ee
Here $G(\eta)$ is the Weyl magnitude for the GW component, introduced earlier in the Brinkmann gauge, but due to gauge invariance for the Weyl entries, we can use the same quantity here as well. (Note we are carrying through this analysis for the `plus' polarization only at the moment, but the results can obviously be extended to the `cross' polarization case as well.)

These equations have the nice feature that we can solve for $m$ and $n$ independently (though in any case any $m$ solution can be turned into an $n$ one, just by flipping $G$ to $-G$). If $G$ and $F$ are harmonic (and if both are present, then with the same frequency $\omega$), then the solutions will be in the form of {\em Mathieu functions}, which we discussed in Section~\ref{sect:harm-sol} above. These do indeed have long-term variations in addition to the expected variations at frequency $\omega$, but as far as we can see this is not due to any division into `background curvature' versus the EM or GW `ripples', as discussed by MTW, rather it seems to be due to a `beating' effect between the driving oscillations and what they induce. Particularly importantly, we can see that the driving component in the GW case is not the ${\beta'}^2$ which appeared in equation (\ref{eqn:Ldd-wave}) and which MTW say represents the energy carried carried by the waves, but the Weyl quantity $G(\eta)$, which in the harmonic case can be both positive and negative, and certainly does not qualify as an energy.

To reinforce that the curvatures go just as $F^2$ and $G$, we finally discuss the Riemann entries again.
It is easy to show that the complete Riemann can be written in the form
\be
\clr(B)=2\pi\clf e_x B e_x \clf -\half G(\eta)e_+ e_x B e_x e_+,
\ee
with the Weyl part corresponding to the second term only. Again, the role of reflections features prominently in this form. It is easy to see from this expression how it satisfies the Einstein equations. If we write $B=a \wdg b$, and take $\da$, then for the Weyl part the answer is proportional to
\be
\da \left(e_+ e_x \left(a\wdg b\right) e_x e_+\right) = \da\left(e_+ e_x \left(ab-a\dt b\right) e_x e_+\right)=0,
\ee
since $\da C a=0$ for any bivector $C$, and the differentiation of $a\dt b$ pulls out $b$ to the left, leaving the $e_+ e_x e_x e_+$ to annihilate. For the non-Weyl part, then noting $\clf e_x=Fe_+$, we get
\be
\da \left(2\pi\clf e_x  \left(ab-a\dt b\right) e_x \clf\right) = 4\pi F^2 e_+ b e_+
\ee
and a further contraction of this via $\db$ yields 0. Noting
\be
4\pi F^2 e_+ b e_+ = 8\pi\left(-\half \clf b \clf \right)
\ee
for any $b$, we see we that this leads to the desired form of SET.

As a final comment on the $m$, $n$ representation, versus $L$ and $\beta$, we note that the controlling equations (\ref{eqn:mn-master}), being linear, are well-adapted to the case where one or both of $m$ and $n$ pass through zero. This will in fact happen eventually for both a pulse-type wave, or harmonic wave, and employing $m$ and $n$ means that this no longer becomes a coordinate singularity, but can be easily dealt with, in similar terms as we have already discussed for the Brinkmann coordinates $\sfx$ and $\sfy$.

\subsection{EM waves in the Brinkmann gauge}

It is worth considering the issue of joint EM and gravitational waves back in the Brinkmann gauge which we used above. The covariant  Faraday tensor is the same as introduced in equation (\ref{eqn:wave-calf}), namely
\be
\clf=F(\eta)\sigma_1\left(1-\sigma_3\right)
\ee
This solves the Maxwell equations, and the SET equation, with $H=G(\eta) f(x,y)$ as used above in Section~\ref{eqn:sect-exact-waves}, becomes
\be
\half G(\eta) \grad^2 f  = 8 \pi F^2(\eta)
\ee
For solutions in cylindrical polars of the form $f(\rho,\phi)=\rho^n \cos(m\phi)$ (note $n$ and $m$ are not related to the variables in the preceding subsection and are intended to be general integers here), then we get the following equation
\be
\rho^n\left(m^2-n^2\right)\cos m\phi +4\rho^2=0
\ee
This then singles out $m=0$, which we can think of as a `breathing mode', and we get the further requirement $n=2$. The solution we want is therefore
\be
H=4 \pi F^2(\eta) \rho^2
\ee
This can then be combined with the gravitational wave solutions found above just by adding. This happens despite everything being exact and (in principle) non-linear, since the relevant controlling equation involves the 2d Laplacian $\grad^2 f(x,y)$, the solutions of which obey a superposition principle. This applies whenever the waves, either gravitational, electromagnetic, or both, are moving in the same direction.

It is interesting to ask whether our geodesic approach used above also applies to combined EM and gravitational waves. This is again non-trivial, since non-linear aspects are involved, in particular the velocity constraint equation (\ref{eqn:v-constraint}) and our general expression for $\sfz$ in (\ref{eqn:z-from-xy}), from which we find both $\sfz$ and $\sft$. The answer, however, is in the affirmative. We can use our results so far for uncharged particle motion in a joint EM/GW wave. The equations we need, replacing the relevant parts of (\ref{eqn:geo-ddots}), are
\be
\begin{gathered}
\ddot{\eta}=0, \quad \ddot{\sfx}=-\dot{\eta}^2\left(G\left(\sfx \cos2\phi_0+\sfy \sin2\phi_0\right)+4\pi F^2 \sfx\right)\\
\ddot{\sfy}=-\dot{\eta}^2\left(G\left(-\sfy \cos2\phi_0+\sfx \sin2\phi_0\right)+4\pi F^2 \sfy\right)\\
\end{gathered}
\label{eqn:geo-ddots-with-EM}
\ee
and then our general expression for $\sfz$ in (\ref{eqn:z-from-xy}) is in fact unchanged, despite the gravitational forces due to the EM wave being included. Finally, we get the time $\sft$, just by adding $\eta=a s$ again. Equation (\ref{eqn:geo-ddots-with-EM}) is nice in showing very directly the different `spin' characteristics of the forces due to the gravitational and electromagnetic waves. For the former, they are clearly spin-2, whereas for the gravitational effect of EM waves, there are no preferred directions, and we can think of this as effectively a `scalar mode', despite the force itself (as it would manifest itself if the particle were charged) having a vector character.

\subsection{Backwards in time particle or photon motion}

An interesting question is whether for strong gravitational waves it is possible to have backwards in time motion for particles or photons as they pass through such a wave. A paper by Penrose, \cite{1965RvMP...37..215P}, says that the spacetime of an exact gravitational plane wave is `physically satisfactory from the point of view of causality' and says specifically that `The fact that it contains no smooth closed timelike or null curves is evident from the form of the metric. On any timelike or null curve not parallel to the propagation world-direction, we can use $u$ as a parameter continuously increasing with time.' Here, Penrose's $u$ is equivalent to our $\eta$, and the idea is presumably that we cannot return to a previous point in the wave's development simply by going forward in time.

\begin{figure}
\centering
\setlength{\lineskip}{\medskipamount}
\subcaptionbox{Massive particle case for a wave with $d=1$. \label{fig:part-time}}{\includegraphics[width=0.49\textwidth]{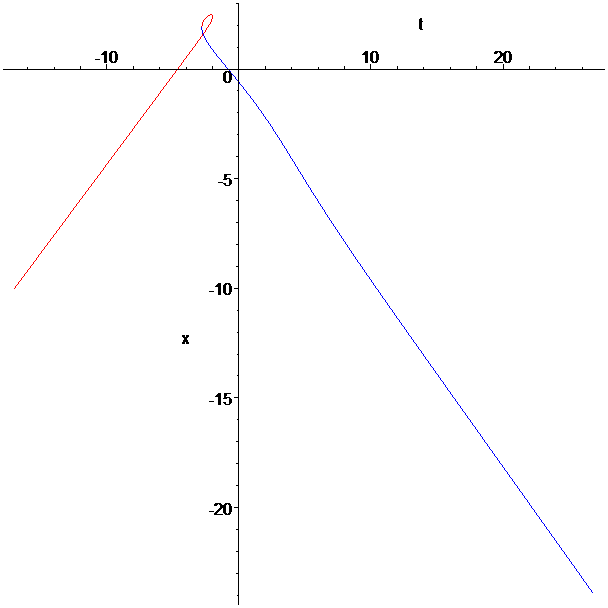}}\hfill
\subcaptionbox{Photon case for a wave with $d=0.3$.\label{fig:phot-time}}{\includegraphics[width=0.49\textwidth]{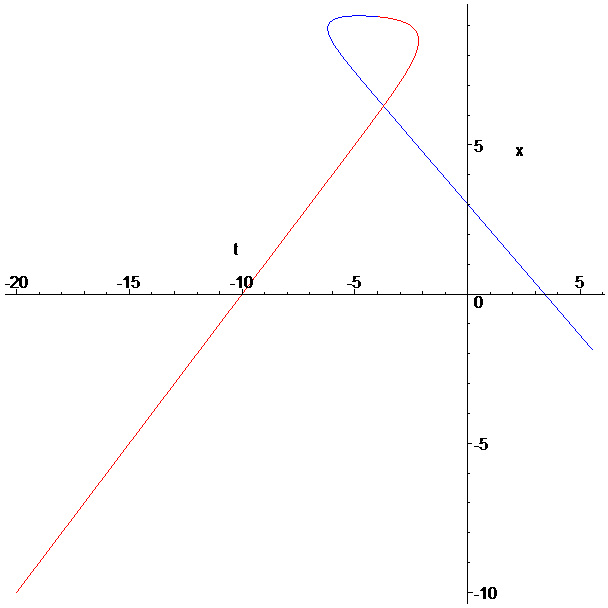}}
\caption{Motion in the $(\sft,\sfx)$ plane for a particle initially moving in the positive $\sfx$ direction, and which encounters an impulsive gravitational wave. Red shows the motion up to $\eta=0$ and blue thereafter. } \label{fig:backwards-in-time}
\end{figure}

\begin{figure}
\begin{center}
\includegraphics[width=0.49\textwidth]{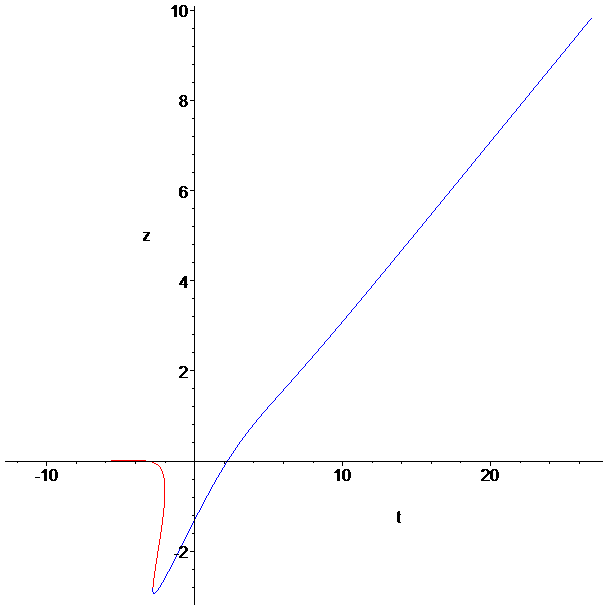}
\includegraphics[width=0.49\textwidth]{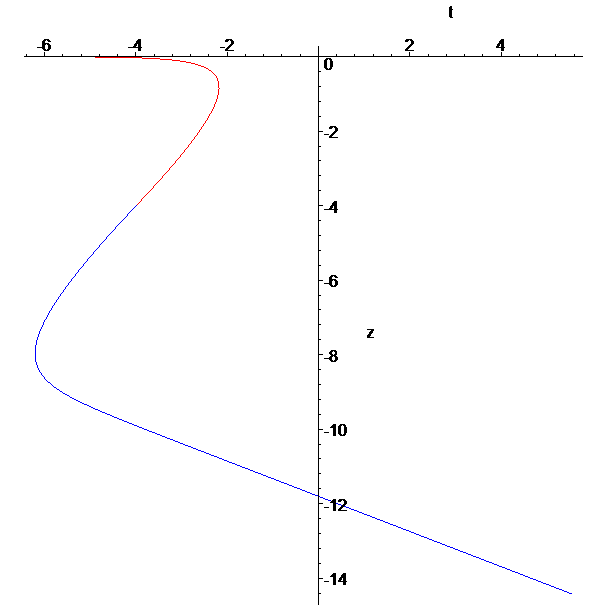}
\caption{Same cases as in Fig.~\ref{fig:backwards-in-time} but showing projections of the trajectories into the $(\sft,\sfz)$ plane.}
\label{fig:tz-plots}
\end{center}
\end{figure}

We agree with this assertion, but there is still the possibility of interesting effects as regards time. Specifically, although $\eta$ is monotonic as a function of $t$, it does not follow that $t$ is monotonic as a function of $\eta$. Thus from the point of view of the motion of a given particle, it is possible for the time associated with it to move backwards for a while, creating an apparent loop in time when projected onto a plane in spacetime. We now give some specific examples of this, both for a massive particle, and a photon. In both cases we consider an impulsive wave of the form given in equation (\ref{eqn:impluse-form}) above, and use the exact solutions available in terms of Bessel functions. These have been given explicitly in (\ref{eqn:imp-exact}) in the massive particle case, and an equivalent expression can also be derived for a photon. Some example results are shown in Fig.~\ref{fig:backwards-in-time},
on the left for a massive particle initially moving in the $\sfx$ direction with a Lorentz boost $\gamma$ factor of 1.7, and on the right for a photon. The wave has amplitude $d=1$ in the left hand case, and $d=0.3$ at the right, so both are very strong non-linear waves. The time constant of the impulse is $\tau=1$ in both cases, and the gravitational wave is timed so that if there were no effects of the wave on the particle/photon motion, it would be centered at $\sfz=0$ when the particle/photon reaches $\sfx=0$. What is plotted is the trajectory of the particles in the $(\sft,\sfx)$ plane with the $\sft$ axis horizontal and the $\sfx$ axis vertical. We see that indeed time can run backwards for a while and that we obtain apparent loops in the $(\sft,\sfx)$ plane. The GW polarization has been chosen so that throughout $\sfy$ is maintained at zero, but one finds that the $\sfz$ component of the motion is {\em different} between the points where the trajectory crosses over itself, i.e.\ the effect of a `loop' is only an apparent one, caused by projecting onto (in this case) the $(\sft,\sfx)$ plane. If we were to look at these motions in 3d, there would be no loop, although there is certainly a period of time running backwards. To verify this, we show in Fig.~\ref{fig:tz-plots}
the projections onto the $(\sft,\sfz)$ plane in the two cases. It is clear that at the points at the beginning and ending of the loops in Fig.~\ref{fig:backwards-in-time}, where the particles return to the same $\sft$ and $\sfx$ values, they do not return to the same $\sfz$ values. This is particularly clear in the photon case, where the development of $\sfz$ with $\sft$ is monotonic. It would be interesting to understand whether there are effects of the backward in time motion that would be apparent physically to a separate observer, perhaps via the particles we have been considering emitting a series of photons, and asking in what order these are received by the observer, but this is the subject of future work.

\section{Summary and Conclusions}

In this contribution, after a review of Geometric Algebra and Gauge Theory Gravity, we have considered gravitational waves from the point of view of these systems, and have found that a relatively simple approach and equations allow description of {\em exact} waves from general relativity. Moreover a hitherto neglected feature of the effect of gravitational waves upon particles as they pass over them, namely a `velocity memory' effect, becomes clear in this approach, and instead of previous notions of collapse of a coordinate system to a point, substitutes physical notions of the induced long term motions of particles.

We have also considered electromagnetic waves, and discussed how combined exact EM and gravitational waves can be simply treated, and how this does not lead to support for a `background curvature' effect of such waves, as advocated by Misner, Thorne and Wheeler. Finally, we looked at the question of whether very strong gravitational waves could induce `backward in time' motion in particles or photons, and found that it could, although exactly what physical effects would be visible in this has yet to be elucidated.

Of interest also in the future for a GA approach to gravitational waves, will be in how these appear in {\em modified gravity} theories. The fact that GTG is based upon the localising the symmetries of rotation and translation, suggests strongly that the extension to local scale-invariance should be looked at as well. In fact the foundations of a complete theory along these lines has now been developed, and written up (so far only in a translation into conventional tensor notation) in Lasenby \& Hobson \cite{2016JMP....57i2505L}. In these approaches, it turns out that {\em torsion} becomes an inescapable component of the propagation of gravitational waves, with new physical effects comparable to the interplay between electric and magnetic fields in the propagation of EM waves. This is an area well worth pursuing further, both for possible new physics, and more generally the opportunities that gravitational wave observations will provide for discriminating between competing theories of gravity.

Overall, we have found a GA and GTG approach very useful for making the mathematics and physics involved in gravitational waves to be understandable to people from a wide variety of backgrounds, needing only small extensions of the GA needed in engineering or other physics areas. We thus recommend this approach more widely, and hope others will find it interesting.


\bibliography{sig_supplementary_references}
\bibliographystyle{birkjourmock}

\end{document}